\newcommand{\revision}[1]{\textcolor{black}{#1}}

\newcommand{\bm}[1]{\mathbf{#1}}
\documentclass[%
 amsmath,amssymb,
 aps,
 prfluids,
]{revtex4-2}

\usepackage{graphicx}
\usepackage{tikz} 
\usepackage{dcolumn}
\usepackage{bm}
\usepackage{float}
\usepackage[normalem]{ulem}
\usepackage{xcolor}

\usepackage{pgfplots}

\usepackage{multirow}
\usepackage{soul}
\setstcolor{red}

\newcommand{\agh}[2][white]{{\sethlcolor{#1}\hl{#2}}}

\begin{document}

\preprint{APS/123-QED}

 \title{Vortex breakdown in a hydro turbine draft tube swirling jet }
 
\author{Artur Gesla
 }

\author{
Eunok Yim  }
 \affiliation{Hydro Energy and Applied fluid Dynamics (HEAD) Laboratory, École Polytechnique Fédérale de Lausanne (EPFL),
   Lausanne, Switzerland}

\date{\today}

\begin{abstract}
The swirling flow in a Francis type hydropower turbine is known to be susceptible to the formation of a large helical structure, commonly referred to as a vortex rope. This vortex rope can be interpreted as an unstable mode associated with vortex breakdown. This perspective is adopted here in a simplified laminar flow setting. The helical vortex rope mode is shown to bifurcate supercritically from an axisymmetric base flow in a Hopf bifurcation within a turbine draft tube. When wall friction effects are neglected, a large recirculation region at the axis can form and a range of subcritical solutions is identified for a flow regime corresponding to partial load of the turbine. The existence of these subcritical solutions promotes the emergence of a hysteresis loop. We further describe a regular dynamics of a formation of recirculation bubble at the axis and its destruction due to the emergence of a helical vortex rope at its periphery. Increasing the axial flow discharge towards the regime corresponding to nominal turbine load leads to an unfolding of the steady solutions branch in a transcritical bifurcation. This bifurcation takes place at finite Reynolds number and complements existing evidence of transcritical bifurcation of the swirling jet flows, previously reported in the inviscid limit.

\end{abstract}

\maketitle






     


\section{Introduction}
The design of a water turbine plays a crucial role in the efficiency of the hydropower generation process. A peculiar hydrodynamic phenomenon, observed specifically inside water turbines of Francis type, manifests itself as an unsteady helical coherent mode inside the draft tube placed right behind the turbine. This mode, commonly known as a \textit{vortex rope}, is first documented in the works of \citet{allen1924comparative} and \citet{rheingans1940power}. The vortex rope commonly arises in the off-design partial load or overload operational regime of the turbine and causes dangerous low frequency (typically few Hz) oscillation that decreases the longevity of the turbine and deteriorates the overall efficiency of power generation.

Efforts to describe and possibly mitigate the effects of the vortex rope have led to many experimental and theoretical, later followed by numerical, investigations. Vortex rope in hydro turbine is usually a turbulent low pressure region of the flow. Air/vaporous bubbles in the flow, either already present in the fluid or generated due to cavitation, allow for experimental visualisations with  examples available in the works of \cite{dorfler1980mathematical,ciocan2007experimental,seidel2014dynamic}.
 Vortex rope is usually composed of a single helical structure that rotates in the direction of the mean azimuthal flow and is chracterized by the azimuthal wavenumber $m=1$. It is now understood that it originates from the presence of residual swirl in the draft tube, as the turbine is not extracting the whole swirl component of the flow in the off-design operating conditions. Most engineering studies focused on \textit{ad hoc} mitigation strategies that include installing additional flow guides inside the draft tube or flow injections in order to suppress rotation (relevant review can be found in \cite{escudier1987confined}). Hydrodynamic origin of the vortex rope has been identified as the fundamental phenomenon of vortex breakdown by \citet{cassidy1970observations}. 

Vortex breakdown can be defined as a `formation of an internal stagnation point on the vortex axis, followed by reversed flow in a region of limited axial extent' \citep{leibovich1978structure} or a more general  `abrupt and drastic change of structure which sometimes occurs in a swirling flow' \citep{benjamin1962theory}. It occurs in many flows, typically characterised by some degree of swirl, and can be either desirable (e.g., the swirling recirculation zones in combustion devices that stabilise the flame and enhance the combustion efficiency) or undesirable effect (as in the flow over delta wings or in the present case). Numerous examples are given by \cite{escudier1987confined}.

The phenomenon manifests itself in laminar flows as a first sign of unsteadiness as well as in compressible supersonic flows \citep{zhang2009topological}. It takes the form of an axisymmetric steady recirculation bubble, a helical or double-helical structure \citep{faler1977disrupted} or a conical sheet \citep{billant1998experimental} among others. Different forms of the vortex breakdown have been reported for analytical swirling jet profiles like Batchelor vortex \citep{delbende1998absolute}, Grabowski vortex \citep{ruith2003three,gallaire2006spiral,vyazmina2009bifurcation} or vortex velocity profiles fitted directly to the experimental data \citep{gallaire2003mode}. The great variety of flows in which the breakdown occurs and forms that it can take, have motivated many interpretations of the phenomenon.

The axisymmetric form of the vortex breakdown \revision{can originate in a base flow evolution }
 as a control parameter, such as the swirl intensity $S$ or the Reynolds number $Re$ (defined later in the text), is varied. It has been shown numerically \citep{tsitverblit1993vortex} that following a branch of steady solutions leads to the formation of first one and then multiple recirculation regions without any axisymmetric bifurcation. This is analogous to the formation of a recirculation region in the flow around a circular cylinder \citep{chen1995bifurcation}, where the formation of a recirculation region is not a bifurcation but simply a smooth evolution of the base flow. In that flow, the first true bifurcation, that leads to the well-known von Kármán vortex street, which breaks the reflection symmetry, is a supercritical Hopf type at $Re\approx40$ depending on the specific configuration (in the present work, the terms sub-/supercritical refer to the existence of nonlinear solution branches around the bifurcation point, and not to the wave speed of disturbances superimposed on the base flow, as in some early studies of vortex breakdown). 
 
\revision{Another, equally relevant scenario for the axisymmetric vortex breakdown includes an axisymmetric bifurcation. \agh{The connection between this loss of stability and the global bifurcation structure of the steady states was established by} \cite{wang1997dynamics}\agh{, who showed that the critical swirl is a point of exchange of stability and that the transition to axisymmetric breakdown can be understood as an evolution from the columnar state to a neighbouring steady state once the incoming swirl reaches or exceeds the critical level.} It has been shown by linear stability analysis \citep{Wang_Rusak_Gong_Liu_2016} and in the direct numerical simulation \citep{Feng_Liu_Rusak_Wang_2018} that the solid body rotation flow with uniform axial velocity is unstable to axisymmetric eigenmodes at a given critical swirl number and finite Reynolds number. Similar scenario for the axisymmetric breakdown was reported in Lamb-Oseen vortex \citep{QiaoPof2025}, where a series of transcritical, saddle-node \agh{and Hopf} bifurcations was identified. Following the branches of solutions involved in these bifurcations led to the formation of a region of decelerated flow at the axis and ultimately flow reversal with the  formation of a recirculation bubble.} 
All these theoretical studies cited demonstrate a high level of 
consistency with DNS data, reinforcing the validity of the transition mechanism.

The helical mode of the breakdown is believed to originate in a bifurcation from a steady solution. This bifurcation has shown to be a supercritical Hopf bifurcation in the experiments of \cite{liang2005experimental,oberleithner2011three}. Both studies report a square root scaling of the amplitude of the helical $m=1$ mode as expected for a supercritical Hopf bifurcation. This is in line with a much earlier inviscid stability analysis of \cite{benjamin1962theory}, claiming that the spiral breakdown originates from an instability of the base flow.

Early computational efforts of analysing the swirling base flow and its stability were limited by the available computing power (\cite{leibovich1978structure} reports only three 2D axisymmetric results available at the time). Numerical stability studies therefore focused on 1D local velocity profiles and the distinction whether the instability was absolute or convective. This was accomplished either by tracking the eigenvalues of the dispersion relation in the complex wavenumber and frequency planes \citep{healey2008inviscid} or by analysing the impulse response in linearised direct numerical simulation (DNS) \citep{delbende1998absolute,gallaire2003mode,gallaire2006spiral}. 
With growing computing power, the analysis was extended to nonparallel flows, one of the first fully 3D direct simulations for vortex breakdown is that of \cite{ruith2003three}. It reports the existence of the recirculation bubble in the steady base flow and then, with growing swirl, an abrupt transition to helical breakdown with $m=1$ or $m=2$. Over some parameter ranges, this transition was accompanied by a periodically growing and bursting axisymmetric recirculation bubble.
The abruptness of the breakdown, sometimes reported in the literature, can be explained by the branch folding scenario and the associated cusp catastrophe proposed by \cite{lopez1994bifurcation}. Plotting a flow observable (either a pointwise quantity or a global norm) over a region of parameter space ($S$, $Re$) reveals that a branch of steady solution can fold back on itself, thereby creating a range of unstable solutions. If $S$ or $Re$ is increased quasi-steadily, as it is often done in experiments or DNS, the system can jump from one stable state to another. This is however observed only in some regions of the parameter space (in the study of \cite{lopez1994bifurcation}, $Re$ has to be above a finite threshold $Re_{cusp}$). 
The bistability described above leads to hysteresis loops, as reported by \cite{billant1998experimental,vyazmina2009bifurcation,meliga2011control} among others with further examples reviewed by \cite{shtern1999collapse}.  

Different computational models have been chosen for simulating the vortex breakdown. A necessarily finite computational domain can be adapted to mimic an open domain by a proper choice of wall boundary conditions, as in \cite{ruith2003three,vyazmina2009bifurcation}. The domain extremities can also be made impermeable with a slip \citep{lopez1994bifurcation} or a no-slip \citep{meliga2011control} boundary condition to simulate a flow in a radially closed domain. Different axial pressure gradients can be imposed by choosing a divergent \citep{sarpkaya1974effect} or a convergent-divergent \citep{lopez1994bifurcation,meliga2011control} geometry. The adverse pressure gradient was shown to greatly promote the onset of breakdown \citep{sarpkaya1974effect}, whereas enforcing no-slip  on the domain walls was shown to delay it \citep{meliga2011control}.

\revision{The question of axial inhomogeneity of the vortex breakdown phenomenon was in particular studied by a series of works started by \cite{wangPof1996a}, who observed unstable axisymmetric breakdown modes in a shearless swirling jet profile previously reported to be stable in the classical (modal ansatz in axial and azimuthal directions) normal mode analysis. The finite pipe length used by these authors together with the Dirichlet inlet conditions have been shown to break the symmetry supposed by the Fourier mode ansatz in the axial direction and this was shown to promote the presence of unstable axisymmetric modes. This observation was later made also for the same solid body rotation with uniform axial flow for non-axisymmetric modes in the linear stability analysis  \citep{Wang_Rusak_Gong_Liu_2016} and in the direct numerical simulation \citep{Feng_Liu_Rusak_Wang_2018}\agh{, with the underlying perturbation energy-transfer mechanism analysed through the Reynolds-Orr equation} \citep{feng2017energy}. Extensions to more realistic Lamb-Oseen vortex profile are provided by \cite{QiaoPof2025} for the axisymmetric modes and \cite{Qiao_Shi_Meng_Wang_Liu_2025} for three dimensional modes.}

The aim of this work is to apply the bifurcation analysis tools and to gain insights from fundamental studies of vortex breakdown to an industrial flow in the draft tube of a Francis hydro turbine. In this study, we use available experimental studies of a reduced-scale model of a real Francis hydro turbine \cite{ciocan2007experimental,susan2006analysis,amini2023upper}, in which a helical vortex rope was identified. In those experiments, the vortex was visualised by cavitation, which expends radially outward with the draft tube's conical shape and rotates in the turbine's direction at about 30\% of the runner rotation speed. Recent global stability analyses based on the turbulent mean flow have found this helical vortex rope corresponds to an instability mode \citep{pasche2017part,muller2021prediction} in reasonable agreement with experimental data.
Additionally, \citep{head_paper_1} investigated local stability properties using the same inlet velocity profile as in the present study. The analysis is essentially 1D, relying on a parallel flow assumption, but they incorporated the measured turbulent viscosity into the stability model. It was found that the flow rate near the best efficiency point is the least unstable, although the $m=0$ and $m=1$ modes still exhibit positive growth rates. For lower flow rate, the $m=0, 1$, and $2$ modes are unstable, and higher azimuthal wavenumbers are more strongly damped as the eddy viscosity increases. \revision{It has to be noted however, that the 1D analysis relying on the parallel flow assumption can yield the result which are not corresponding to the 2D results are demonstrated by \cite{wangPof1996a} for a vortex breakdown in a solid body rotation or \citep{QiaoPof2025} for the Lamb-Oseen vortex. Additionally, in the present configuration neither the flow itself nor the geometry are strictly invariant in the axial direction. Therefore, the comparison between classical modal 1D and 2D analysis has to be taken with even more caution. }

The saturation mechanism following the helical instability yielding vortex rope and whether the associated bifurcation is sub- or supercritical are yet to be clarified. 
 It is furthermore not clear whether the presence of turbulence is essential for the vortex rope formation in this flow or whether qualitative insight can be obtained from the study of a laminar flow. 
 The current work uses a realistic inflow condition based on the three-vortex model proposed by \cite{susan2010analysis} and their axisymmetric computational domain in order to analyse the onset and saturation mechanism of the vortex rope. The inflow model is fitted to the experimental measurements and the computational domain is constructed to resemble a true draft tube geometry. The onset of vortex rope formation and its saturation mechanism are studied using linear stability analysis and direct numerical simulations (DNS).
  The present work focuses on the bifurcation scenario in the laminar regime and constitutes an initial step towards extending the investigation to the turbulent case in future work. Nevertheless, even in the laminar regime, we show that the vortex rope can be identified and its rich dynamics including sub- and supercritical bifurcations, bistability and hysteresis loops clearly connect it to the broader class of vortex breakdown phenomena. In addition, we consider two different side wall boundary conditions (no-slip and free slip) to assess the impact of wall boundary layers on the rope dynamics. 

The article is structured as follows. Section \ref{sec2} describes the flow domain and the numerical tools used. Section \ref{sec:baseflow} presents the base flow and its linear stability. Section \ref{sec:nonlin} analyses the flow dynamics obtained from time integration and summarises observations made for the case of no-slip boundary conditions on the outer wall. This constraint is relaxed in section \ref{sec:slip} where a different bifurcation scenario is identified for a free slip wall. Different operating regimes of the turbine are analysed in section \ref{sec:dop}. Finally, the summary and outlooks are presented in section \ref{sec:summ}. 

\section{Numerical model} \label{sec2}
\begin{figure}[]
    \centering
    \includegraphics[width=0.7\linewidth]{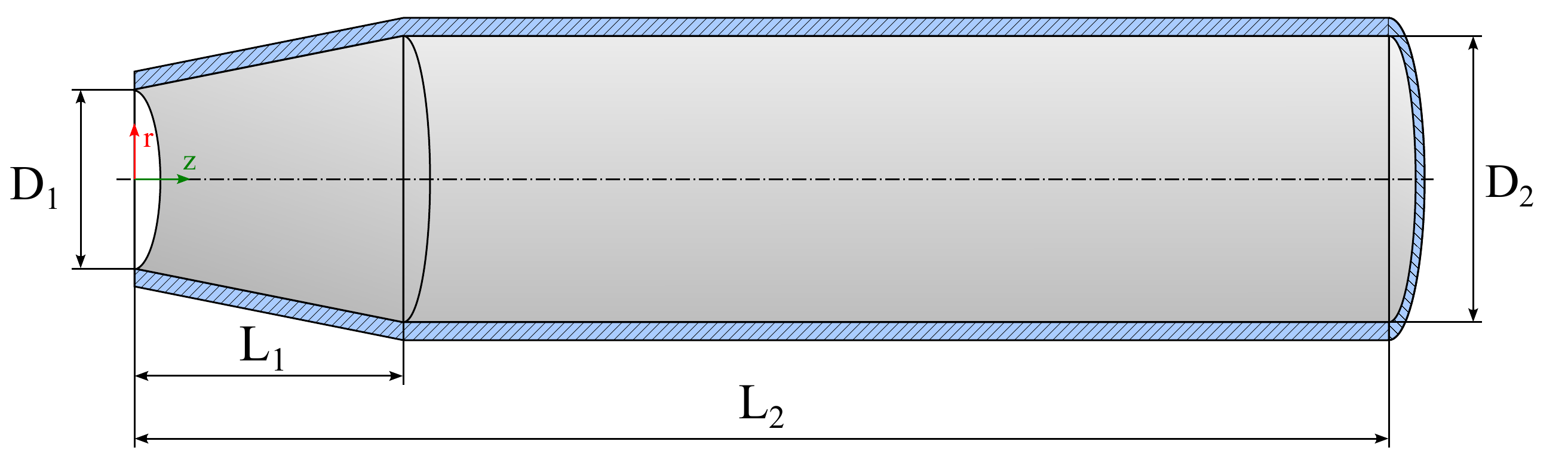}
    \caption{Domain composed of a conical diffuser and a straight cylindrical section \citep{susan2010analysis}. Flow enters the domain on the left and exists on the right. $R_1=D_1/2=1.063$, $R_2=D_2/2=1.6$, $L_1=3.58$, $L_2=15$.  \revision{The opening angle of the conical diffuser part is $\arctan((R_2-R_1)/L_1)\approx 8.5\deg$.}
    }
    \label{fig1}
\end{figure}

Flow domain is axisymmetric and composed of a conical diffuser and a cylindrical pipe following \cite{susan2010analysis} (see figure \ref{fig1}). Flow enters the domain in the left where the inflow velocity profile ($\mathbf u_{\,inlet}$) is imposed following the experimental fit proposed by \cite{susan2006analysis}:
\begin{gather} \label{inletprof}
U_r=0, \\
U_{\theta}=\Omega_0 r + \frac{\Omega_1 r_1^2}{r}\left[1-\exp(-r^2/r_1^2)\right]+ \frac{\Omega_2 r_2^2}{r}\left[1-\exp(-r^2/r_2^2)\right], \\
U_z=U_0+U_1\exp(-r^2/r_1^2)+U_2\exp(-r^2/r_2^2). \label{inletprof2}
\end{gather}

\begin{figure}
\centering
    \includegraphics[width=0.39\textwidth]{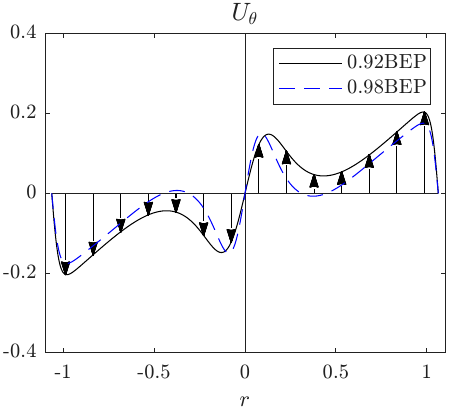}
    \hspace{20pt}
    \includegraphics[width=0.39\textwidth]{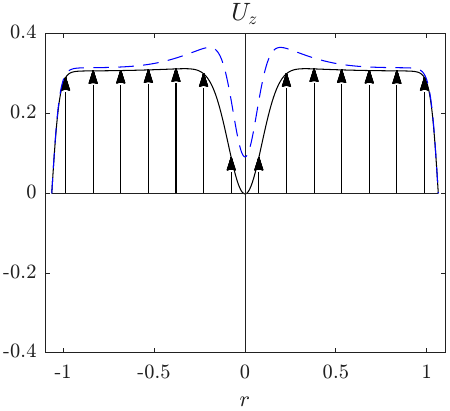}
    \caption{Inlet velocity profile corresponding to the 0.92 and 0.98 best efficiency point 
    (black solid and blue dashed line respectively). Parameter values are specified in table \ref{tab1}.
Radial velocity $U_r=0$. Additional exponential regularisation is applied near the wall at $r=1.063$ as discussed in the text.
}
    \label{inlet}
\end{figure}

The values of nondimensional numerical parameters are given in table \ref{tab1}. Reference length- and timescales of the system are chosen to be $R_0$ and $1/\Omega$, where $\Omega$ is the angular frequency of the turbine runner of radius $R_0$. The resulting velocity scale is $R_0\Omega$ and the Reynolds number is $Re=R_0^2\Omega/\nu$ with $\nu$ the kinematic viscosity. Dimensional values of these parameters associated to a Francis turbine in \cite{susan2006analysis} are $R_0=20\ \text{cm}$ and $\Omega=1000\ \text{rpm}$ resulting in $Re\approx4\times10^6$. Two turbine operating points are characterised by the discharge coefficient $\varphi$ which is proportional to the axial flux $\int U_z r {\mathrm{d}}r$ and by the fraction $\varphi/\varphi_{BEP}$, where $\varphi_{BEP}=0.368$ corresponds to the turbine best efficiency point (BEP). In order to avoid strong shear near the solid wall, the profiles \eqref{inletprof}-\eqref{inletprof2} are regularised \revision{(multiplied)} with a function 
$f(r)=(1-\exp(\varepsilon(r-r_{\max})/r_{\max}))/(1+\exp(\varepsilon(r-r_{\max})/r_{\max}))$ with $r_{\max}=1.063$ being the inlet outer radius and $\varepsilon=50$. Resulting inlet velocity profiles are shown in figure \ref{inlet} with the 0.92BEP profile characterised by larger swirl component and lower axial flux compared to the 0.98BEP counterpart. 

\begin{table}[]
\centering
\begin{tabular}{cc|cccccccc}
$\varphi$&BEP  & $\Omega_0$ & $\Omega_1$ &  $\Omega_2$&  
$U_0$& $U_1$& $U_2$& $r_1$& $r_2$\\
\hline \hline
0.34 &  0.92 & 0.31765 & -0.62888& 2.2545& 0.30697 &0.01056& -0.31889 &0.46643& 0.13051   \\ 
0.36 & 0.98 &  0.26675 &-0.79994& 3.3512& 0.31501 &0.07324& -0.29672 &0.36339 &0.09304          \\
\end{tabular}
\caption{Inflow profile parameters extracted from \cite{susan2006analysis}. The discharge coefficient $\varphi$ is defined in \cite{susan2006analysis} and is reproduced here for convenience. }
\label{tab1}
\end{table}

\begin{figure}
\centering
    \includegraphics[width=0.99\textwidth]{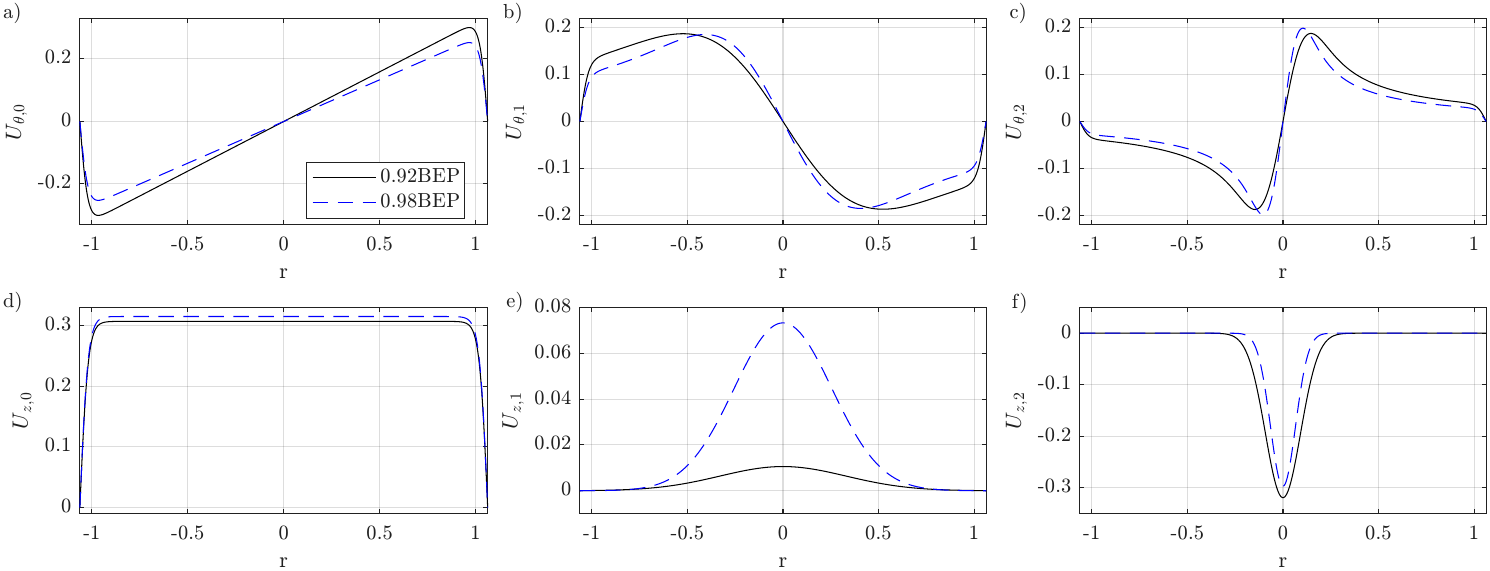}
    \caption{\revision{Decomposition of the inlet profiles defined in equations \eqref{inletprof}-\eqref{inletprof2} on three elementary components : a solid body rotation with a uniform axial flux and two counter rotating Batchelor vortices for two turbine operating points (cf. table \ref{tab1} for the values of numerical parameters).}
}
    \label{inletdeco}
\end{figure}

\revision{Inlet velocity profiles presented in figure \ref{inlet} are additionally decomposed on three elementary vortices constituting the profiles defined in equations \eqref{inletprof}-\eqref{inletprof2} : i) a solid body rotation with angular velocity $\Omega_0$ and uniform axial velocity $U_0$ (figure \ref{inletdeco} a,d), ii) a Batchelor vortex with a core radius $r_1$ rotating at an angular velocity $\Omega_1$ and an axial jet of amplitude $U_1$ (figure \ref{inletdeco} b,e) and iii) a second, counter rotating with respect to the first one, Batchelor vortex with a core radius $r_2$ rotating at an angular velocity $\Omega_2$ and an axial wake of amplitude $U_2$ (figure \ref{inletdeco} c,f). Values of the numerical parameters can be compared between table \ref{tab1} and figure \ref{inletdeco}. The same regularisation function, as defined in the previous paragraph, was applied to the elementary velocity profiles.
}

Flow is governed by incompressible Navier-Stokes equations:
\begin{eqnarray} 
\frac{\partial {\bm u}}{\partial t} + ({\bm u}\cdot {\bm \nabla}){\bm u}& =&-{\bm \nabla}p + \frac{1}{Re}{\bm \nabla}^2 {\bm u}, \label{nseq}\\
{\bm \nabla} \cdot {\bm u}&=&0. \label{contieq}
\end{eqnarray}
Here,  the pressure $p$ is normalized with $\rho\,\Omega^2R_0^2$.
Steady state solution and its stability are analysed 
by computing a weak formulation of \eqref{nseq}-\eqref{contieq} and discretising using finite element method as employed in the FreeFem++ software \citep{freefem}. Details on finite element treatment specific to fluid problems can be found in \cite{pironneau1989finite}. Planar mesh in $(r,z)$ plane consist of 23 594 triangular Taylor-Hood elements with P2--P1 polynomial orders corresponding to velocity and pressure (discussion on other element choices can be found in \cite{ern2004theory}). Sensitivity of selected results to the mesh size is presented in the appendix A. Steady base flow solution $\left[ \bm U_b,p_b\right]$ of the nonlinear governing equations is identified with Newton--Raphson iterative procedure. Boundary conditions are of Dirichlet type on the inlet, either no-slip or free slip on the solid wall and $u_r=0$, $u_{\theta}=0$, ${\partial u_z}/{\partial r}=0$ at the axis. Dirichlet conditions are enforced in FreeFem++ with a penalty method and a free slip condition is enforced using the method of Lagrange multipliers, with details given in the appendix B. Outlet condition is $ ({Re}^{-1}\bm \nabla \bm u -p \bm I)\cdot \bm n = 0 $
 where $\bm{n}$ is the outward unit normal and $\bm{I}$ is the identity tensor. This condition is imposed naturally by the finite element formulation, since integration by parts of the pressure and viscous terms yields the corresponding boundary term. Linear stability of the base flow to infinitesimal perturbations is determined by linearising the governing equation
\begin{eqnarray} 
\label{lsa1}
\frac{\partial { \bm{{u}}}}{\partial t} + ({\bm U_b}\cdot {\bm \nabla}){\bm{{u}}}+ ({\bm{{u}}}\cdot {\bm \nabla}){\bm U_b} & =&-{\bm \nabla} p + \frac{1}{Re}{\bm \nabla}^2 {\bm{{u}}}, \\
{\bm \nabla} \cdot {\bm{{u}}}&=&0, \label{lsa2}
\end{eqnarray}
and assuming the form 
\begin{eqnarray} 
\bm u(r,\theta,z,t)= \bm {\hat{u}}(r,z) \exp(\mathrm{i}m\theta+\lambda t), \\
 p(r,\theta,z,t)=  {\hat{p}}(r,z) \exp(\mathrm{i}m\theta+\lambda t),
\end{eqnarray} 
where $m$ is the azimuthal wavenumber, $\lambda$ the complex frequency with homogeneous version of the base flow boundary conditions apart from the axis $r=0$ where the boundary condition depends on the value of $m$:
\begin{equation}
	\begin{cases} 
		\hat{u}_r = \hat{u}_\theta = 0,\ \ \frac{\partial \hat{u}_z}{\partial r}=0  & m\text{ is even} ,\\
				\frac{\partial \hat{u}_r}{\partial r} = \frac{\partial \hat{u}_{\theta}}{\partial r} = 0,\ \ \hat{u}_z=0  & m\text{ is odd} .\\
	\end{cases}
\end{equation}
These conditions are enforced by the cylindrical geometry, with the same formulation used, for example, in \cite{openpipeflow}.

Finite element discretisation of \eqref{lsa1}-\eqref{lsa2} yields a generalised eigenvalue problem of the form 
\begin{equation}
\label{genevp}
 \bm A \bm{\hat{q}} = \lambda \bm B \bm{\hat{q}},
\end{equation}
where $\bm{\hat{q}} = [\bm{\hat{u}},{\hat{p}}] $.
A subset of eigenvalues of \eqref{genevp} nearest to a complex shift $\sigma$ is identified using the shift and invert procedure with the ARPACK library called directly from FreeFem++. Eigenvectors are normalised so that $\int \bm{\hat{u}}^*\cdot\bm{\hat{u}}\;r\,\mathrm{d}r\,\mathrm{d}z=1$ where $^*$ denotes complex conjugate.

Time integration of the Navier-Stokes equations is also \agh{performed} with Nek5000 spectral element code \citep{nek5000}. Mesh of either 1536 (free slip simulations) or 6240 (no-slip simulation) hexahedral spectral elements is used with polynomial order 7 and 5 in each spatial direction of the three dimensional domain for the discretisation of velocity and pressure respectively. Dealiasing with polynomial of order 11 is used. No explicit stabilizing filters are used. Backward differentiation formula of second order (BDF2) is used to advance the equations in time with Courant number $\text{CFL}<0.3$. Boundary conditions are of Dirichlet type at the inlet, no-slip or slip at the outer solid wall and an open outflow condition (details in \cite{nek5000}) at the outlet. Full stress formulation is used in order to apply a slip condition at the outer solid wall.

\section{Base flow and its stability} \label{sec:baseflow}
\begin{figure}[]
    \centering
        \includegraphics[width=0.8\linewidth]{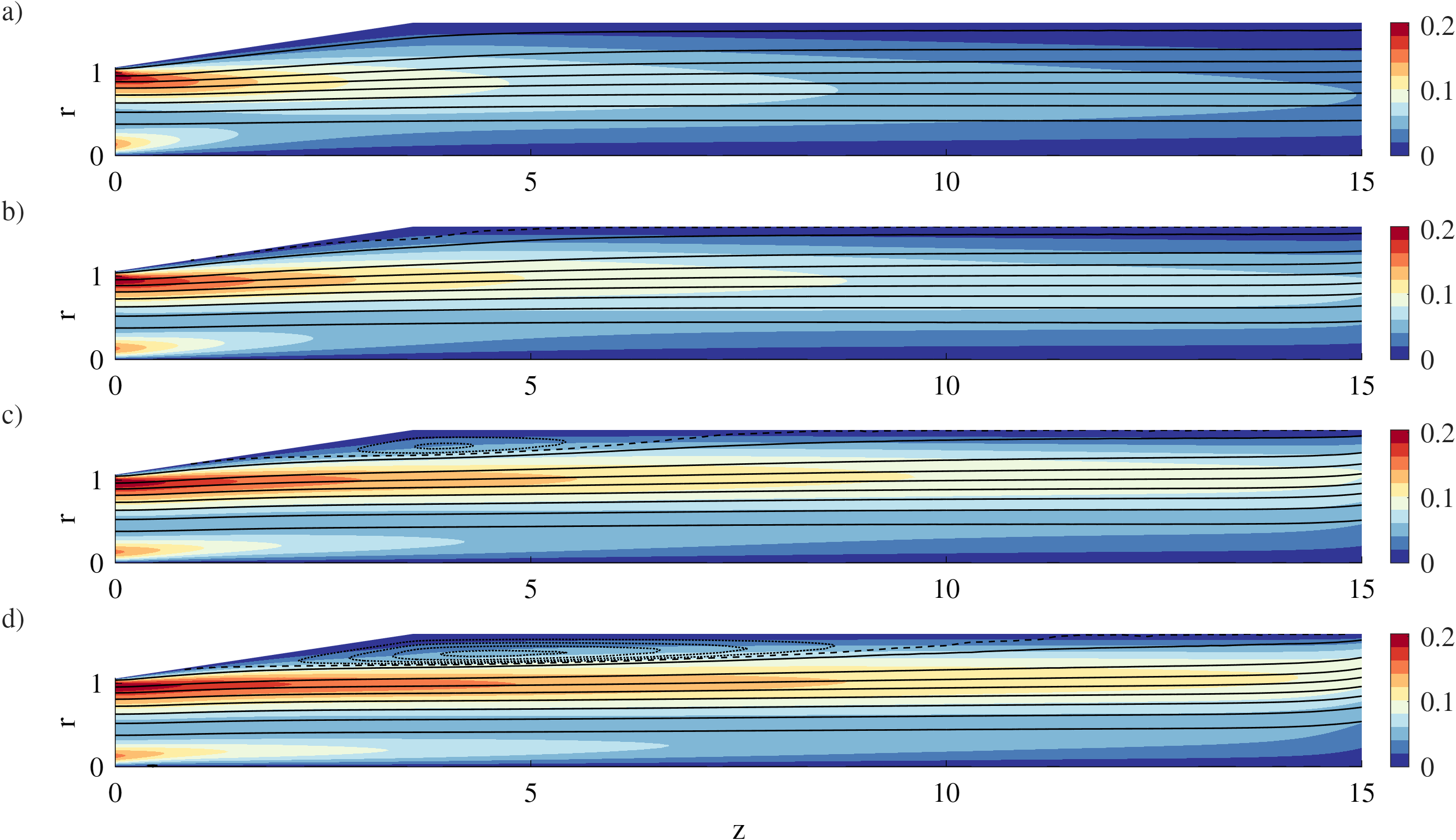}
    \caption{Azimuthal component of the base flow (colorscale) for $Re=500$ (a), $Re=1000$ (b), $Re=2000$ (c) and $Re=3000$ (d). Overlapped isocontour of streamfunction $\psi$ ranges from 0 to 0.16 with increment 0.02 (solid line), equals $\psi_1=0.162$ (dashed line), and ranges from $\psi_1+10^{-3}$ to the maximal $\psi$ value with increment $10^{-3}$ (dotted line).}
     \label{fig2}
\end{figure}

Steady axisymmetric solution of the \eqref{nseq}-\eqref{contieq} together with a no-slip condition at the outer wall is computed for $Re=500,\ 1000,\ 2000,$ and $3000$. For these $Re$ the Newton--Raphson algorithm converges to the solution starting from a homogeneous uniform initial guess so that no continuation procedure is necessary. Streamlines, which are the constant values of the streamfunction $\psi$ of the inplane velocity $(u_r,u_z)$, is computed in post-processing by solving 
\begin{equation}\label{eq:psi}
\frac{\partial^2 \psi}{\partial r^2}+\frac{\partial^2 \psi}{\partial z^2}=u_z+r\left(\frac{\partial u_z}{\partial r}-\frac{\partial u_r}{\partial z}\right),
\end{equation} 
with the $\psi=\psi_0=0$ at the axis and ${\partial \psi}/{\partial r}=r\,u_z$, ${\partial \psi}/{\partial z}=-r\,u_r$ at the other boundaries in FreeFem++. Additionally, the value of $\psi$ at the outer wall can be computed by integrating the inflow profile $\psi_1=\int_0^{{r_{\max}}} u_z\,r\,dr=0.162$.
 Azimuthal velocity $u_\theta$ and the contours of $\psi$ are \agh{shown} in figure \ref{fig2}. For all $Re$ base flow consists of the flow swirling inside the domain with decreasing azimuthal velocity with increasing $z$. \revision{Due to the action of viscosity the inlet profile, which is not a an exact solution of the governing equations, is evolving with increasing $z$. Velocity gradient in the axial direction is stronger for higher viscosity (lower Re), as can be seen in figure \ref{fig2}.} Due to the weak formulation outlet condition the streamlines exit the domain in a direction which is not perpendicular to the outlet. At $Re=1000$, the contour $\psi_1$ detaches from the wall which indicates the presence of a recirculation bubble in the corner where the diffuser and pipe regions of the domain meet. This bubble grows in size with $Re$ and its interior is visualised with isocontour of $\psi>\psi_1$.  

Linear stability of the base flow is computed by solving the eigenproblem \eqref{genevp} for selected nonnegative values of $m$. Scanning the spectrum for eigenvalues $\lambda$ with positive real part reveals that the first mode to become linearly unstable is $m=1$. Precise value of the critical $Re$ is determined with a secant method to be $Re_c=2340.86$ with the angular frequency of the mode $\lambda_i=-1.262$. \revision{This critical $Re$ almost does not depend on the length of the domain $L_2$ ($L_2=20$ : $Re_c=2342.28$, $L_2=30$ : $Re_c=2341.87$). This is because the conical section of the draft tube and the highest gradients of velocity are both localised near the domain inlet. As these are possibly the factors responsible for the presence of the unstable mode, making the domain longer has a negligible effect on the critical threshold. A counter example in given in \cite{Qiao_Shi_Meng_Wang_Liu_2025} who introduces an additional forcing term to maintain the inlet profile through the straight pipe domain. Axial invariance of the base flow and the domain geometry result in a considerable sensitivity of the stability prediction with respect to the domain length in their study (see their figure 27). }

\begin{figure}
    \centering
    \includegraphics[width=\linewidth]{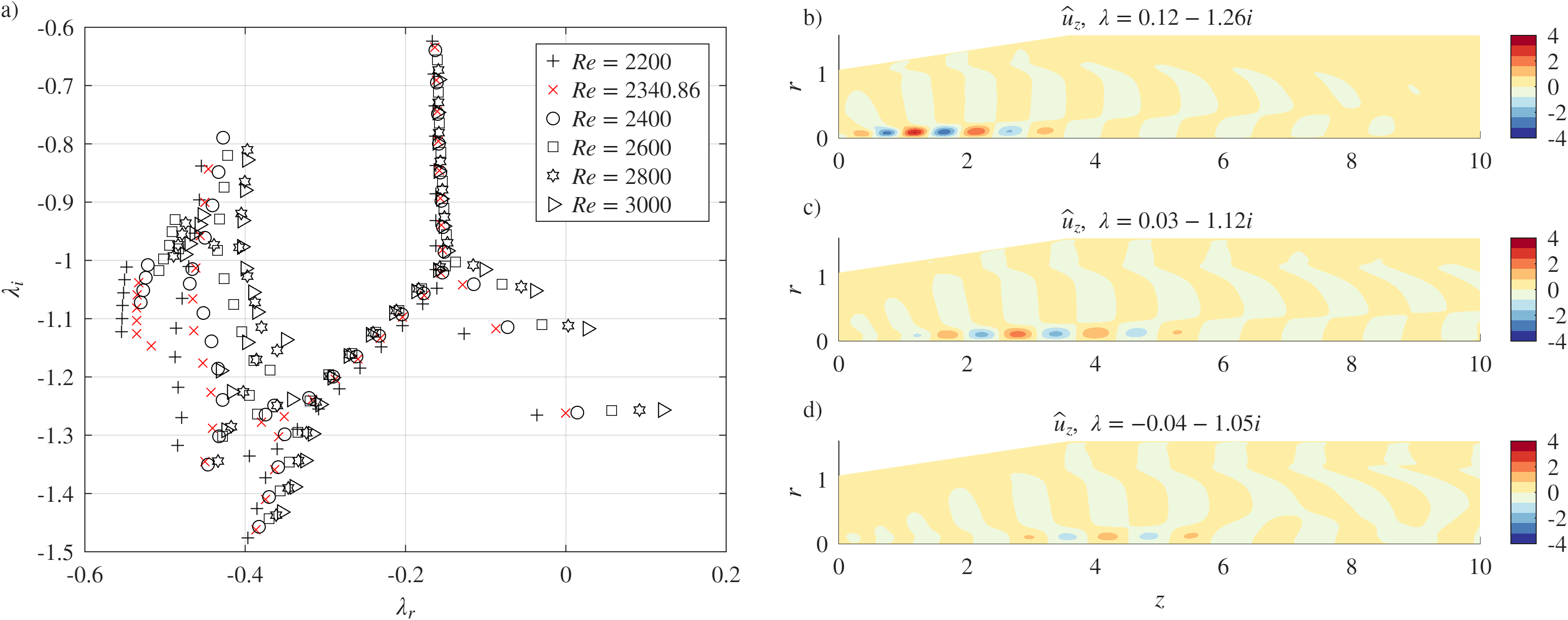}
    \caption{Spectrum of the linearised Navier-Stokes operator for the azimuthal wavenumber $m=1$ (a). Subset of 40 eigenvalues closest to the shift $\sigma=-1.1i$ is shown. The critical Reynolds number is $Re_c\approx2340.86$. The real part of the axial velocity $\hat{u}_z$ of three eigenvectors of leading growth rate at $Re=3000$ is shown in the panels (b-d). Amplitude of the eigenvector is determined by the chosen normalisation.}
     \label{fig:spec}
\end{figure}

Spectrum of the linearised Navier-Stokes operator and the axial velocity $\hat{u}_z$ of selected eigenvectors is shown in figure \ref{fig:spec}. For the leading eigenvalues the corresponding eigenvector consists of structures localised at the symmetry axis of the model. As these eigenvectors are characterised by the azimuthal wavenumber $m=1$, the eigenvector corresponds to a monohelical structure. This results compares favourably with various experimental observations -- draft tube vortex rope is mostly a $m=1$ mode concentrated at the axis and localised right at the runner outlet in the diffuser part of the draft tube. Direct comparison is of course limited since the present result is obtained for a low $Re$ laminar flow and the vortex rope is usually a fully turbulent phenomenon. Additional differences between the unstable laminar modes and the experiment observation include the spatial structure of the rope and its frequency. Vortex rope is usually characterised by frequency which is a fraction of the runner frequency (typically $0.2-0.4\,\Omega$ reported by \cite{amini2023upper}) and a single helix following a conical shape (distance between the axis and the helix grows with downstream direction). Modes shown in figure \ref{fig:spec} are rather characterised by frequency around $1.1-1.3\,\Omega$ and the mode is concentrated within a cylindrical region near the axis. These differences are additionally addressed further in the discussion part. Instability of the base flow at finite $Re$ leads to a departure of the dynamics from the base flow for $Re>Re_c$ and can be followed by a saturation to a nonlinear state. This scenario is analysed in the next section. 

\revision{\agh{The experimentally observed vortex rope can, in specific turbine operating points, take the form different from the $m=1$ helical shape. In particular, the vortex rope in the turbine partial load close to the best efficiency point can take a double helical ($m=2$) form and the vortex rope in overload regime can be an axisymmetric ($m=0$) pulsating structure}} \cite{Muri02}. \revision {\agh{Additionally, the laminar vortex breakdown is often marked by the competition between $m=0,1,2$ modes (e.g.}} \cite{ruith2003three}\revision{\agh{). These observations motivate the search for unstable modes at $m\neq1$.}}

\begin{figure}
    \centering
    \includegraphics[width=\linewidth]{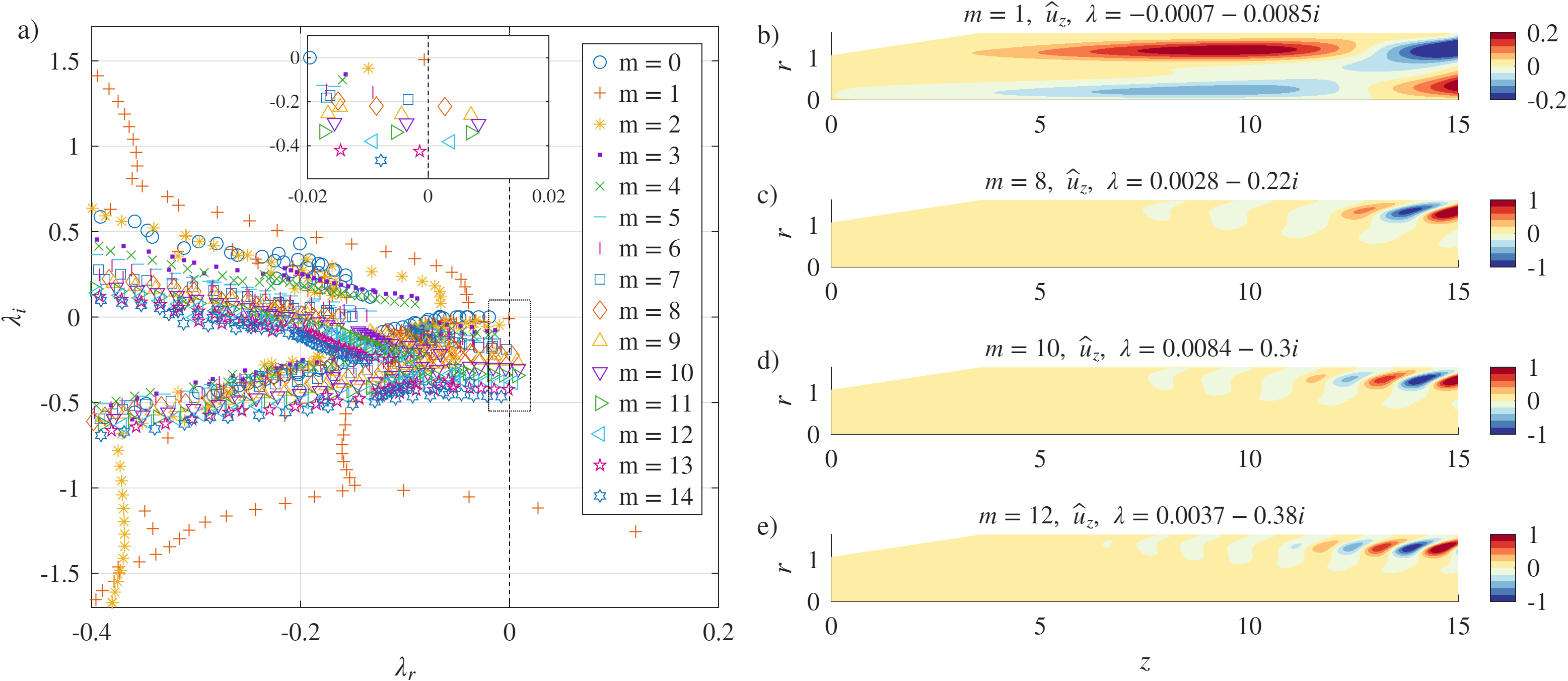}
    \caption{\revision{\agh{Spectrum of the linearised Navier-Stokes operator for the azimuthal wavenumbers $m\in[0,14]$ for $Re=3000$ (a). Inset in the panel a) is a zoom of the region of the spectrum where modes $m\in[8,12]$ are unstable. The real part of the axial velocity $\hat{u}_z$ of selected eigenvectors is shown in the panels (b-e).}}}
     \label{fig:higherm}
\end{figure}

\revision{\agh{ A sweep of eigenvalue searches is performed for $m\in[0,14]$ for a series of shift values $\sigma\in(-2i,2i)$ with a procedure described in section}} \ref{sec2}\revision{\agh{ at $Re=3000$. The resulting spectrum is shown in figure}} \ref{fig:higherm}\revision{\agh{. All of the $m<8$, apart from the leading $m=1$ helical mode, are stable. A range of unstable helical modes include $m\in[8,12]$ with $m=10$ marked by the highest growth rate. The selected eigenvectors corresponding to these unstable modes are shown in the panels c) - e) in the same figure. These eigenmodes are characterised by a wave growing in the boundary layer that forms on the no-slip walls of the cylindrical domain. They are conceptually reminiscent of Görtler vortices but this analogy is not verified here in detail. As these modes seem to be disconnected from the vortex breakdown phenomenon they are noted here for completeness but not analysed further. Importantly, $m=0$ and $m=2$ modes are stable in the current analysis and it is expected that the vortex breakdown phenomenon will be clearly governed by $m=1$ helical mode.  }}

{\agh{ 
Until this point, all of the presented results concern the inflow profile corresponding to the 0.92BEP conditions. In this regime of turbine operation, the vortex rope has been clearly observed experimentally. 
As 1.0BEP is the most stable operation condition, it is interesting to investigate the influence of changing the velocity profile (effectively increasing BEP) on the computations presented above.
}}

{\agh{ 
A velocity profile corresponding to 0.98BEP is plotted against its 0.92BEP counterpart in figure}}
 \ref{inlet}{\agh{ . Higher BEP profile is characterised by lower swirl (lower $U_{\theta}$). Presence of the vortex rope is connected to the presence of the residual swirl in the flow so the higher BEP the weaker the rope should be. Higher BEP profile is also marked by a higher total axial flux (higher $U_z$). This is because the discharge through the turbine increases with the relative flow rate \%BEP, so the flow rate at 0.98BEP is higher than at 0.92BEP.}}
\agh{To investigate the effect of the changing velocity profile on the linear stability result, a continuous change of the profile is proposed between the profiles at 0.92BEP and 0.98BEP} \revision{such that the parameters ($\Omega_0,\Omega_1,\Omega_2,U_0,U_1,U_2,r_1,r_2$) listed in table \ref{tab1} are linearly interpolated between the operating points for each parameter $p_i$} :

\begin{equation} \label{eq:interp}
p_i=p_{i,\,0.92BEP}(1-\alpha)+\alpha\;p_{i,\,0.98BEP},
\end{equation}
\agh{
where $\alpha$ is the interpolation parameter, $\alpha=0$ corresponds to 0.92BEP inlet profile and $\alpha=1$ to the 0.98BEP profile.}

\begin{figure}
    \centering
    \includegraphics[width=\linewidth]{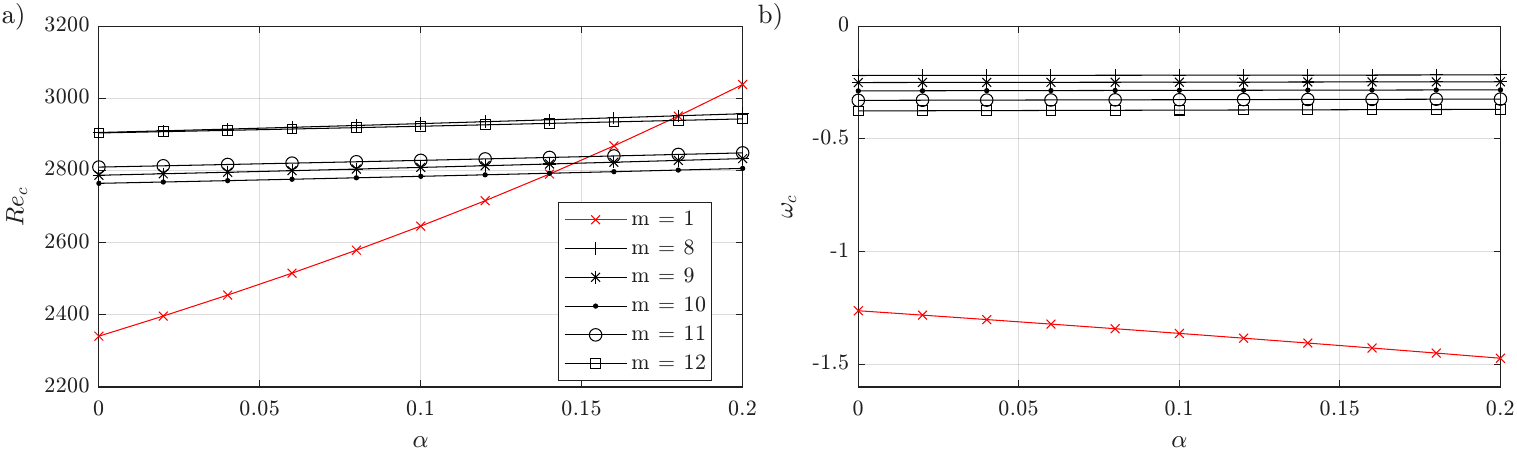}
    \caption{\revision{\agh{Evolution of the threshold $Re_c$ as a function of interpolation parameter $\alpha$ (see eq.}} \eqref{eq:interp}) \revision{\agh{for a series of $m$ (a) and the angular frequency of the eigenmode at criticality (b). In both plots the $m=1$ is highly sensitive to changing the turbine operating point (effectively $\alpha$) while $m\in[8,12]$ are almost insensitive to this change. This supports the observation that it is only the $m=1$ which is connected to the vortex breakdown phenomenon.}}}
     \label{fig:rechighm}
\end{figure}

\revision{\agh{The evolution of the critical $Re$ for the unstable modes identified in figure }}\ref{fig:higherm}\revision{\agh{ as a function of the interpolation parameter $\alpha$ is shown in figure }}\ref{fig:rechighm}\revision{\agh{. $Re_c$ is identified for each $\alpha$ and $m$ with a secant method. With $\alpha$ growing from 0, the $Re_{c, m=1}$ grows substantially, and it is not the most unstable mode any more at $\alpha=0.2$. On the other hand, the $Re_c$ for modes $m\in[8,12]$ is largely unaffected by the inflow profile change. This further supports the observation that these modes are boundary layer modes disconnected from the vortex breakdown phenomenon. An analogous observation can be made concerning the mode angular frequency $\omega_c$ at criticality. It is additionally noted that the $m=1$ mode stabilises for growing $\alpha$ (correspondingly BEP growing towards 1), and this is in line with experimental evidence of the helical vortex rope.}}

\section{Nonlinear saturated states}\label{sec:nonlin}
Departure from the linearly unstable base flow identified in section \ref{sec:baseflow} is analysed by performing a direct numerical simulation of the governing equations \eqref{nseq}-\eqref{contieq} at a series of $Re$. For the generation of an appropriate initial condition the base flow solution is recomputed with FreeFem++ on a fine grid consisting of 275 516 triangles  for each $Re$ and mapped as a axisymmetric initial state to Nek5000. An energy observable is used to monitor the departure of the dynamics from the base flow:
\begin{equation} 
\label{enpert}
E_{pert}(t)=\frac{1}{2}\int_0^4\int_0^{2\pi}\int_0^{0.5}u_r'^2+u_{\theta}'^2+u_z'^2\; r\,{\mathrm{d}}r\,{\mathrm{d}}\theta\,{\mathrm{d}}z,
\end{equation}
with 
$\bm{u}'(t)=\bm{u}(t)-\bm{U}_b$
 being the difference between the instantaneous velocity and the one mapped to the initial field. 
  In the case of perfect correspondence between the spatial discretisation used in two codes $E_{pert}\rightarrow0$ as the solution approaches steady state in the time integration. A small residual level of $E_{pert}$ corresponds to a difference between steady fields obtained with planar finite element discretisation (FreeFem++) and three dimensional spectral element one (Nek5000). The observable $E_{pert}$ is defined is a cylindrical subdomain $z\in[0,4]$ and $r\in[0,0.5]$ in order to focus only on the saturation amplitude of the unstable mode within this subdomain. 

\begin{figure}[]
    \centering
    \includegraphics[width=0.7\linewidth]{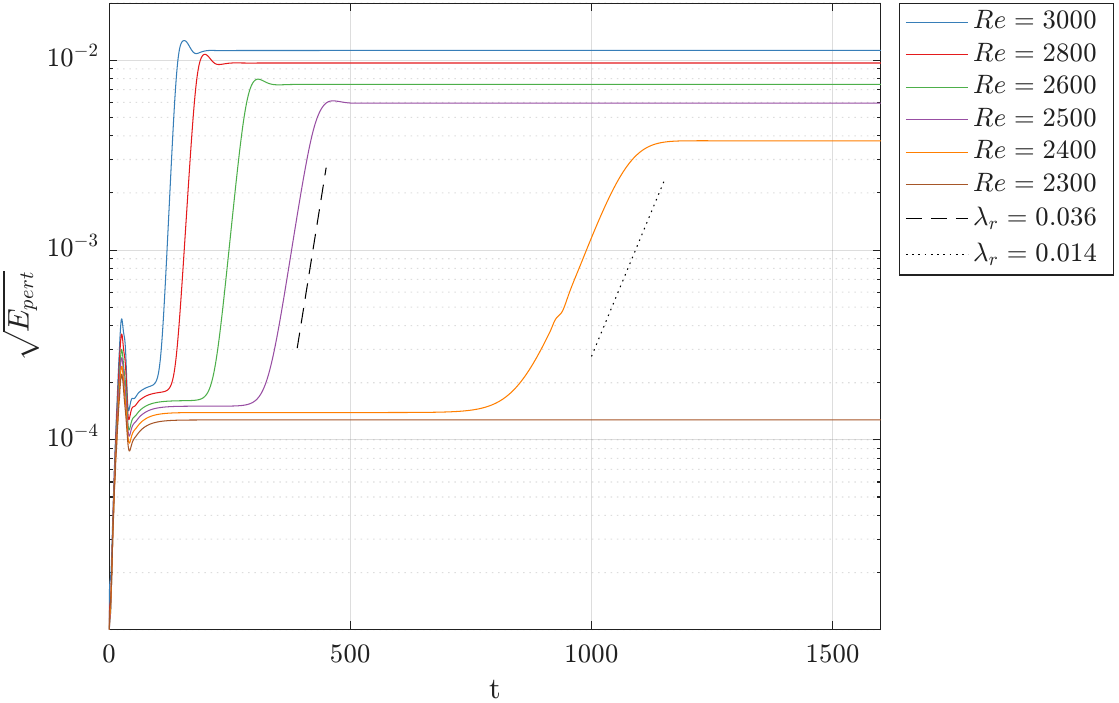}
    \caption{Time series of perturbation energy observable \eqref{enpert} upon Nek5000 time integration of the steady state solution obtained with FreeFem++. The observable remains at a low residual level for $Re<Re_c$ and saturates nonlinearly for $Re>Re_c$. Rate of the exponential growth is compared with the results from linear stability analysis for $Re=2400$ and $2500$. }
     \label{fig:epertt}
\end{figure}

Time series of $\sqrt{E_{pert}}$ for $Re$ from 2300 to 3000 is shown in figure \ref{fig:epertt}. At $Re=2300<Re_c$, the observable converges to $O(10^{-4})$, a low level corresponding to a stable steady solution. At $Re>Re_c$ departure from the steady solution is marked by an exponential growth of the observable. The rate of exponential growth agrees with the prediction obtained from linear stability analysis performed in section \ref{sec:baseflow}. A saturation to a nonlinear state follows and the amplitude of this saturated state corresponds to the saturation level of $\sqrt{E_{pert}}$.  

\begin{figure}
    \centering
        \raisebox{80pt}{a)}
                \includegraphics[width=0.85\linewidth]{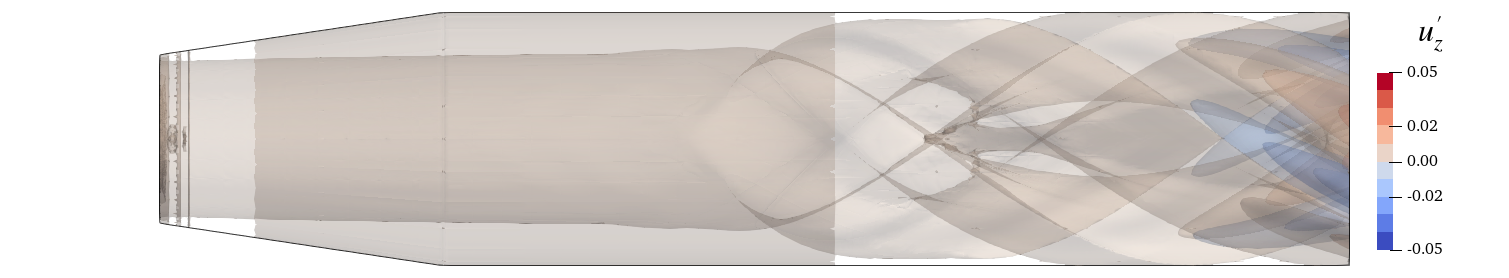}
        \newline
            \raisebox{80pt}{b)}
                \includegraphics[width=0.85\linewidth]{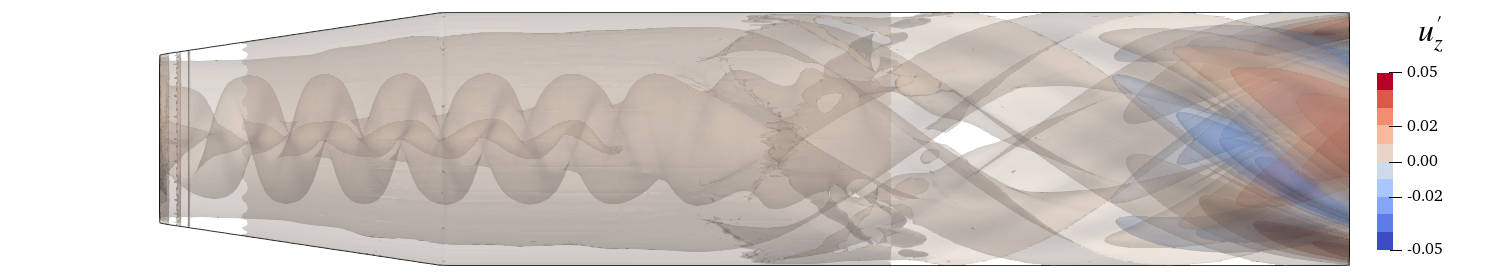}
        \newline
            \raisebox{80pt}{c)}
                                \includegraphics[width=0.85\linewidth]{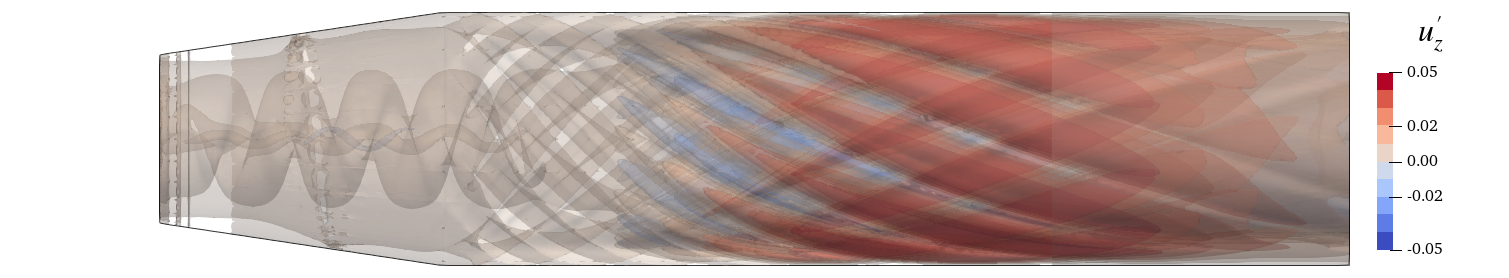}
                \newline
    \caption{Isosurfaces of axial velocity perturbation {${u}_z'={u}_z-{U}_{b,z}$} at $t=1000$ for $Re=2300$ (a), $Re=2500$ (b) and $Re=3000$ (c). Additional video presenting the whole time integration is provided as a supplementary material \citep{suppmat}.}
     \label{fig:endti}
\end{figure}

Snapshots of the saturated solution at $t=1000$ are visualised for $Re=2300,\,2500$ and $3000$ in figure \ref{fig:endti}. For $Re>Re_c$, a clear vortex rope structure corresponding to the linearly unstable mode is identified. Exponential growth and a nonlinear saturation process of this mode corresponds  to the evolution of the observable in figure \ref{fig:epertt}. Additionally, even for $Re=2300<Re_c$, a mode of approximately $m=4-8$ appears near the outlet and, for $Re=3000$, near the domain center. 
Series of modes concentrated at the outer wall near the outlet is identified in the spectrum for even azimuthal wavenumbers but all of theses modes are stable (e.g. for $Re=2500$, the least stable modes are $\lambda_{m=2}=-0.015-0.047i$, $\lambda_{m=4}=-0.022-0.095i$, $\lambda_{m=6}=-0.021-0.143i$, $\lambda_{m=8}=-0.015-0.202i$).
It was confirmed that the presence and the spatial location of this mode is not affected by doubling the domain length or by refining the mesh by approximately a factor of four in the original domain. A series of probes was also placed along the radius at $z=14$ and the signal was found to be steady at $Re=2300$, disordered but with a spectrum detecting the vortex rope frequency at $Re=2500$ and chaotic (with no preferred frequency) at $Re=3000$. It is therefore assumed that this mode is connected to the receptivity of the boundary layer that forms at the wall of the domain and it is neither a linearly unstable mode \revision{\agh{(at least for $Re=2300$ and $Re=2500$, cf. thresholds for instability of $m\in[8,12]$ modes in figure}} \ref{fig:rechighm}\revision{\agh{)}} nor a mode originating from the outlet boundary condition. This is supported by the fact that a series of \agh{stable} modes of small $|\lambda_r|$ is present in the spectrum and this is known to cause large values of gain due to nonnormality in shear flows (e.g. \cite{Cerqueira_Sipp_2014,Gesla24}). Its origin could be linked to the mesh shape (see fig. \ref{fig:meshes}) which introduces a small $m=4$ perturbation. However, as this mode does not affect the formation and saturation process of the vortex rope, the investigation of its precise origin is left aside in the present work and the mode is neglected in the following discussion. In order to focus only on the $m=1$ vortex rope, we restrict the evaluation of the observable $E_{pert}$ to a subdomain that represents only a fraction of the full computational domain.

\begin{figure}
    \centering
    \includegraphics[width=\linewidth]{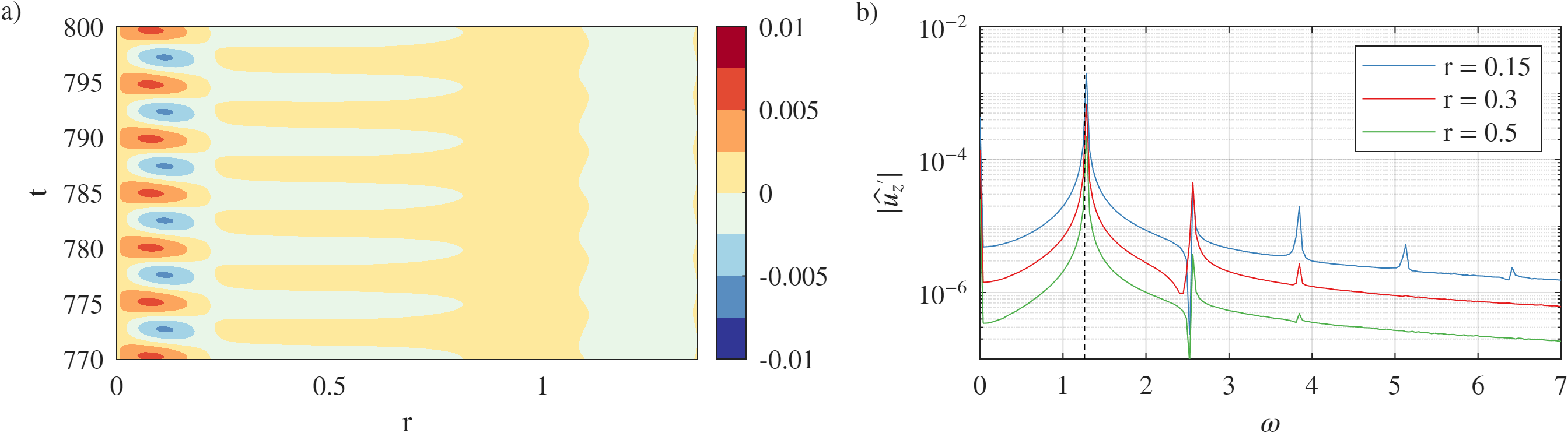}
        \includegraphics[width=\linewidth]{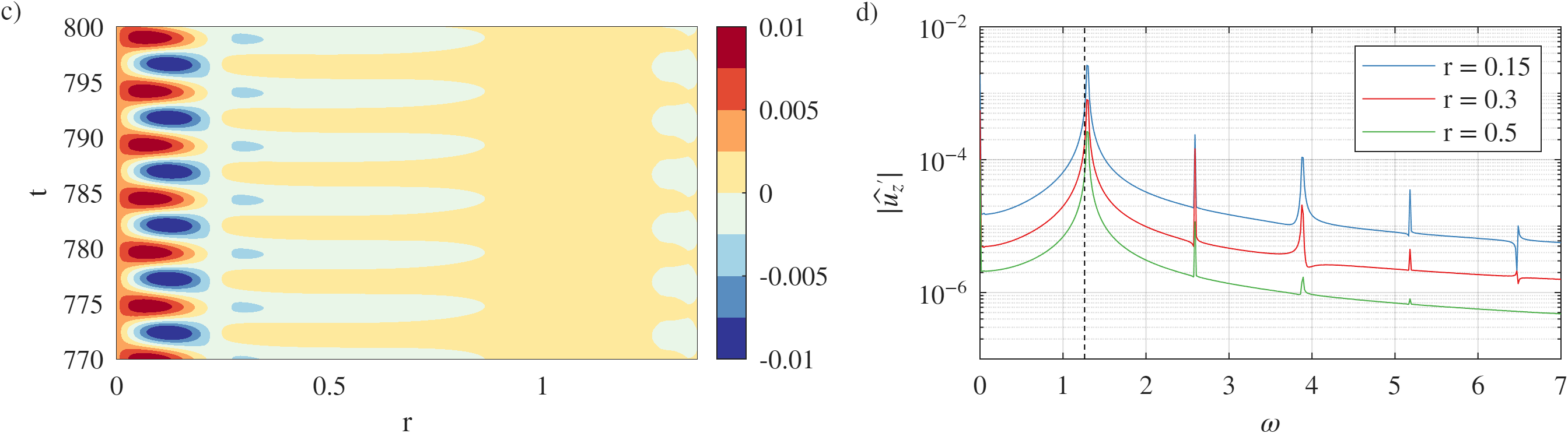}
    \caption{Spacetime diagram for the perturbation axial velocity $u'_z$ at $z=2$ for (a) $Re=2500$, (c) $Re=3000$ and corresponding spectrum amplitude $|\hat{u}_z'|$ of the probes signal (b, d) at three selected radii. Periodic dynamics in the spacetime diagram correspond to the distinctive peaks in the spectrum. Angular frequency of the most unstable mode at the corresponding $Re$ is marked with the dashed line in panels (b, d). Discrete Fourier transform is calculated from the signal of length of 400 time units yielding the spectrum resolution $\text{d}\omega=0.016$.}
     \label{fig:std}
\end{figure}

Saturated vortex rope is a structure rotating periodically around the model symmetry axis. Spacetime diagram of the axial velocity perturbation signal is plotted in figure \ref{fig:std} for $Re=2500$ and $3000$. Amplitude of the oscillation is, as expected, larger for $Re=3000$. Oscillation frequency is measured by performing a discrete Fourier transform of the signal at \agh{three} distinct radial locations. Main peak of the frequency spectrum is in a good agreement with the angular frequency identified in the linear stability analysis $\omega\approx1.2$. A series of harmonics is also identified in the signal and their amplitudes are decreasing with increasing $\omega$. A more pronounced presence of higher harmonics for growing $Re$ is associated with the departure from a purely sinusoidal oscillation originating at $Re_c$. It is also noted that even if two eigenvalues are unstable at $Re=3000$ in the spectrum in figure \ref{fig:spec}, only one frequency is seen in the time signal. This is because the solution saturates at the branch originating at $Re_c$ and is not influenced by a second, possible unstable at the onset, branch of oscillatory solutions associated to a second eigenvalue becoming unstable at $Re>Re_c$.

\begin{figure}
    \centering
    \includegraphics[width=\linewidth]{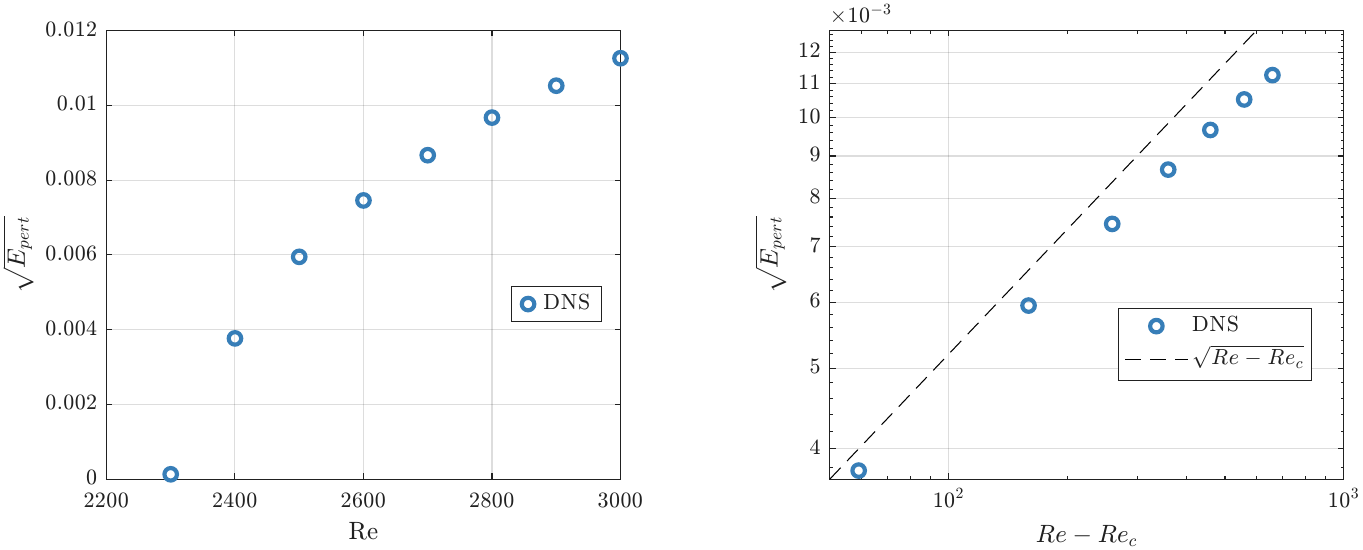}
    \caption{Saturation amplitude of the vortex rope as a function of $Re$. Square root scaling, characteristic of a supercritical Hopf bifurcation, is seen when the data is plotted in the log-log scale. }
     \label{fig:satamp}
\end{figure}

As seen in figure \ref{fig:epertt}, the saturation amplitude of the unstable mode grows with $Re$. Figure \ref{fig:satamp} shows that this dependence is approximately $\sqrt{E_{pert}}\propto\sqrt{Re-Re_c}$, at least close to the instability threshold. This scaling is characteristic for a supercritical Hopf bifurcation. This type of bifurcation is common in fluid dynamics and is also observed, for example, in flow past a circular cylinder \citep{barkley2006linear} and in a flow in a differentially heated cavity \citep{xin2001linear}.

\subsection*{Discussion of the no-slip wall results}

The scenario for vortex rope formation in the present configuration is relatively simple. For $Re<Re_c$, the flow is axisymmetric and stable to both infinitesimal and finite amplitude perturbations. At $Re=Re_c$, a supercritical Hopf bifurcation takes place and the flow loses linear stability to a pair of complex conjugate eigenvalues associated with a monohelical $m=1$ mode. At this point, a small limit cycle is born and its amplitude scales like $\sqrt{Re-Re_c}$ in the bifurcation vicinity. The limit cycle corresponds to a vortex rope rotating around the symmetry axis with one dominating frequency which is almost unchanged by the nonlinear saturation process and correspond to the imaginary part of the complex eigenvalue yielding the branch. 

Even though the present result is obtained for a laminar flow a certain qualitative comparison with the experimental evidence can be drawn. As in experiment, the saturated state identified here is a large self-sustained coherent structure characterised by well established frequency. The azimuthal wavenumber and the axial location of the mode also agree with the experimental observation. 

Notable differences with experiment include the radial form of the mode (the rope is more cylindrical than conical in shape) and the frequency of the mode. The recirculation regions present in the base flow (figure \ref{fig2}) are located at the diffuser-pipe junction, whereas the presence of stalled or recirculation regions is rather observed at the axis in the previous vortex rope investigations \citep{susan2009three}. An additional difficulty is the presence of a spurious high azimuthal wavenumber mode at the wall which is connected with the receptivity of the boundary layer in this work. In what follows, a certain strategy to circumvent some of these problems is proposed.  

\section{Influence of the wall boundary conditions}\label{sec:slip}

The differences outlined in the previous section could be connected to the no-slip condition imposed on the outer radius of the domain. Its presence is resulting in a substantial boundary layer at the solid wall. This boundary layer would be naturally much thinner in a fully turbulent case. The presence of the boundary layer, together with the recirculation regions that form, can be imagined to squash the vortex rope and prevent it from taking a more conical shape. The mode which is concentrated at small radius could correspond to a frequency rise if some form of angular momentum would to be conserved. Eliminating the no-slip condition would naturally eliminate any modes resulting from receptivity of the boundary layer. The reasoning presented above leads to the analysis of the flow with the same inlet and outlet conditions but with a free slip condition imposed on the solid wall. The results are reported below.

\subsection*{Steady base flow solution}

A free slip condition (no shear stress, no permeability) is imposed on the outer radius of the domain. This condition is imposed with Lagrange multipliers method in the context of finite elements with details discussed in the appendix B. \revision{The same inflow profile and regularisation close to the wall is used as in section \ref{sec2}. Even though the free slip condition employed in this section does not require the velocity to vanish at the wall and, in principle, regularising the inflow profiles could have been discarded, it is retained for easy comparison with the results from previous section. Any difference will therefore be a \agh{consequence} of different wall boundary condition and not of different inlet profile. Additional \agh{sensitivity} evaluation of the selected results to the width of the regularisation is given in appendix A.}

Figure \ref{fig:slip} shows the base flow obtained with the Newton method for a series of $Re$. As in figure \ref{fig2}, the azimuthal velocity and streamlines following the definition in \eqref{eq:psi} are plotted. The values of $\psi$ at the wall is not changed as the inflow profile remained unaltered. A notable difference with respect to figure \ref{fig2} is that now the recirculation bubble is located at the axis. The configuration with no bubble (fig. \ref{fig:slip}a), point where bubble starts to appear (fig. \ref{fig:slip}b) and a bubble occupying a considerable part of the domain (fig. \ref{fig:slip}e) are possible. In panel fig. \ref{fig:slip}e even a second recirculation bubble is present near the outlet. A configuration of multiple recirculation bubbles was reported previously for the vortex breakdown case in the cylindrical container \citep{escudier1988vortex} and for a free vortex \citep{viola2016resonance}.

\begin{figure}
    \centering
        \includegraphics[width=0.8\linewidth]{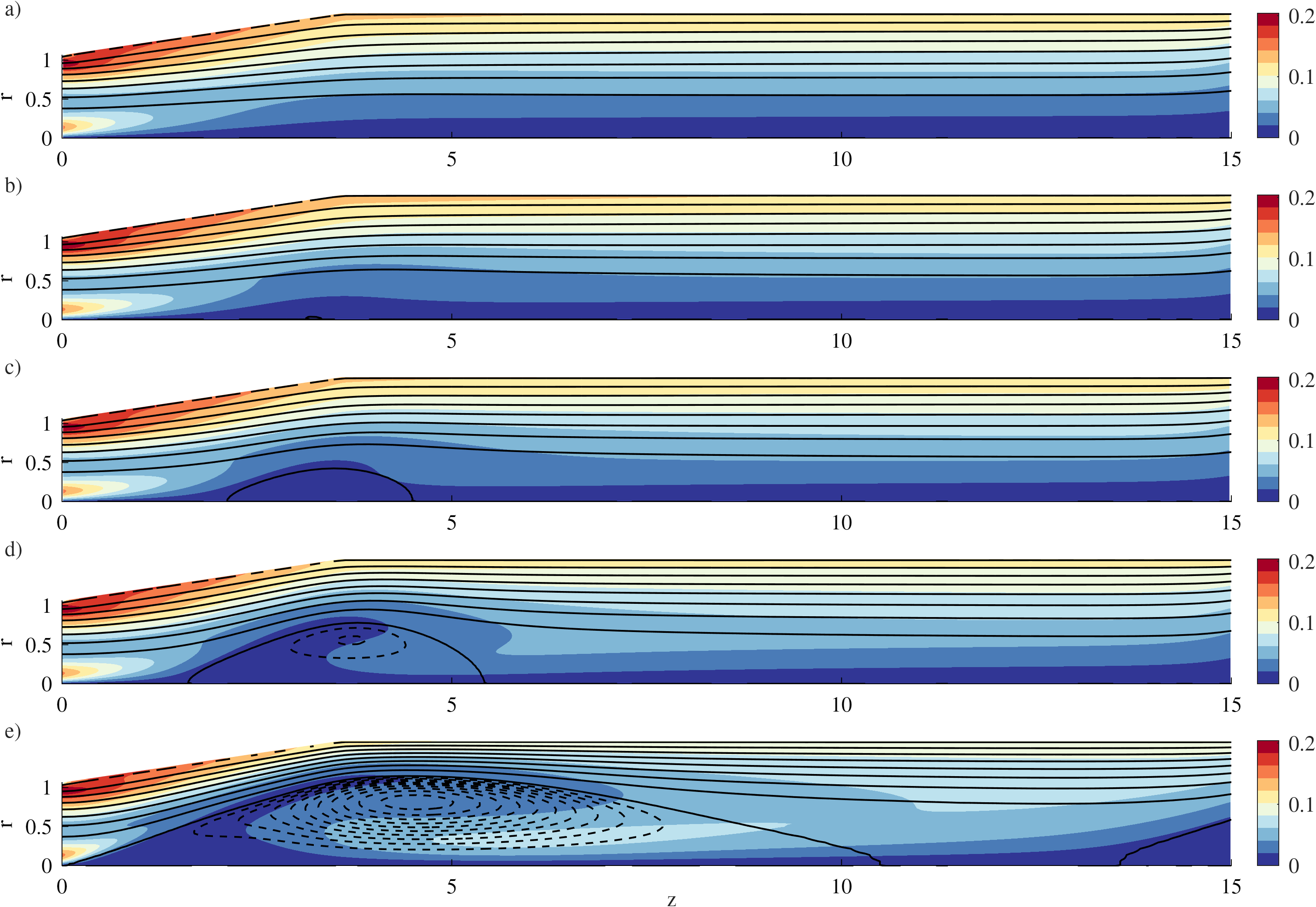}
    \caption{Azimuthal component of the base flow (colorscale)  
     for $Re=491.9$ (a), $Re=813.3$ (b), $Re=749.7$ (c), $Re=677.2$ (d) and $Re=1000.7$ (e). Nonmonotonic increase in $Re$ is caused by following a branch of solutions with an arclength continuation procedure seen in figure \ref{fig:branchal0}. The isocontours of $\psi=0:0.02:0.16$ are plotted in a solid line and of $\psi=\psi_{min}:3\times10^{-3}:-3\times10^{-3}$ in a dashed line.
     }
                  \label{fig:slip}
\end{figure}

\begin{figure}
\includegraphics[width=\textwidth]{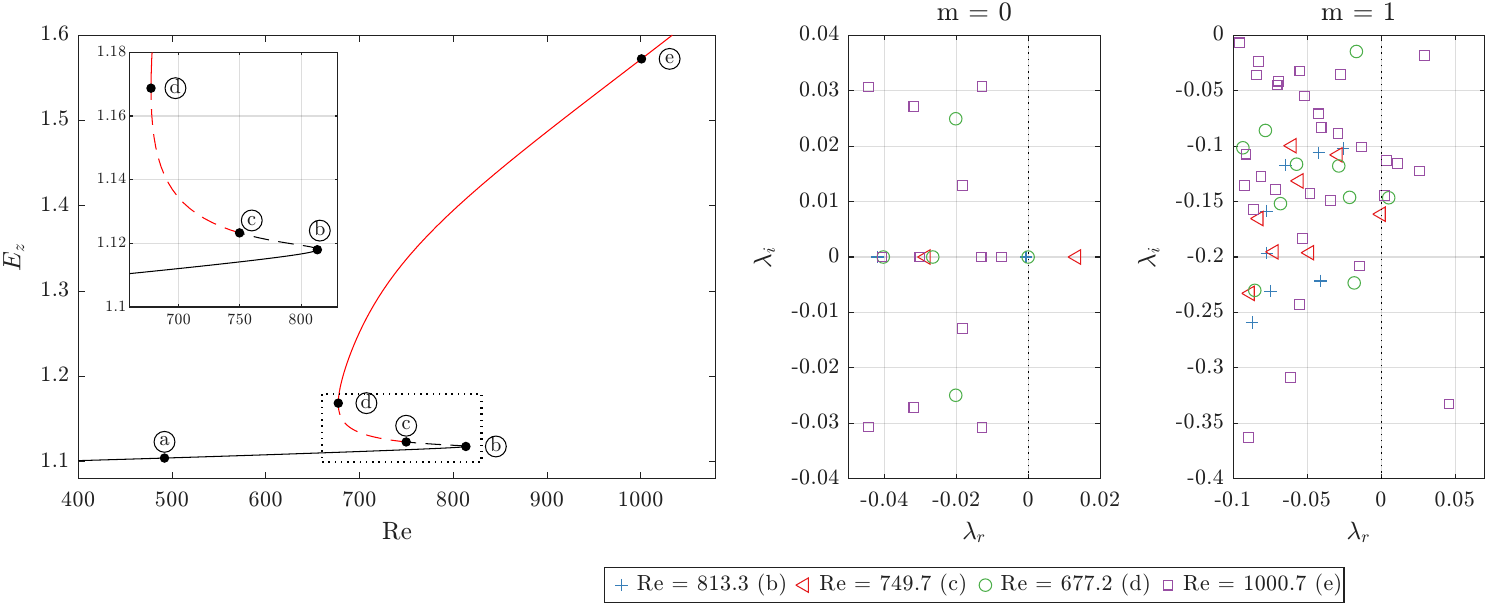}
\caption{Branch of steady solution obtained with an arclength continuation procedure (left) together with its stability at selected points (two panels in the right). Labels \textcircled{a}-\textcircled{e} correspond to the base flow fields plotted in figure \ref{fig:slip} a)-e). Stability to $m=0$ modes is marked with solid (stable) or dashed (unstable) line. Stability to $m=1$ modes is marked with black (stable) and red (unstable) colour. Inset corresponds to the region bordered by the dotted line which zooms on two folds present on the branch.}
\label{fig:branchal0}
\end{figure}

Base flow fields represented in figure \ref{fig:slip} correspond to a pseudo--arclength continuation along the branch of steady solutions identified at low $Re$. Newton method initiated with a zero uniform initial guess converges to this branch when $Re$ is lower than $Re$ at the saddle-node bifurcation point ($Re_{SN}\approx813.3$). \revision{\agh{This $Re_{SN}$ is almost independent of the length of the domain $L_2$ ($L_2=20$ : $Re_{SN}=812.42$, $L_2=30$ : $Re_{SN}=812.49$). }} For $Re>Re_{SN}$, the Newton method diverges even with a initial guess constructed from a solution at a neighbouring $Re$ and an arclength continuation method is necessary to advance along the branch. A standard algorithm of arclength continuation \citep{allgower2012numerical} is used  with the arclength step adapted as a function of a convergence rate of the solution. Details are discussed in the appendix C. The identified branch is traced in figure \ref{fig:branchal0} with an energy observable based on an axial velocity over the entire computation domain 
\begin{equation} \label{eq:ez}
E_{z}=\frac{1}{2}\int_0^{15}\int_0^{2\pi}\int_0^{R(z)}u_z^2\; r\,{\mathrm{d}}r\,{\mathrm{d}}\theta\,{\mathrm{d}}z.
\end{equation}

\begin{figure}
\includegraphics[width=0.9\textwidth]{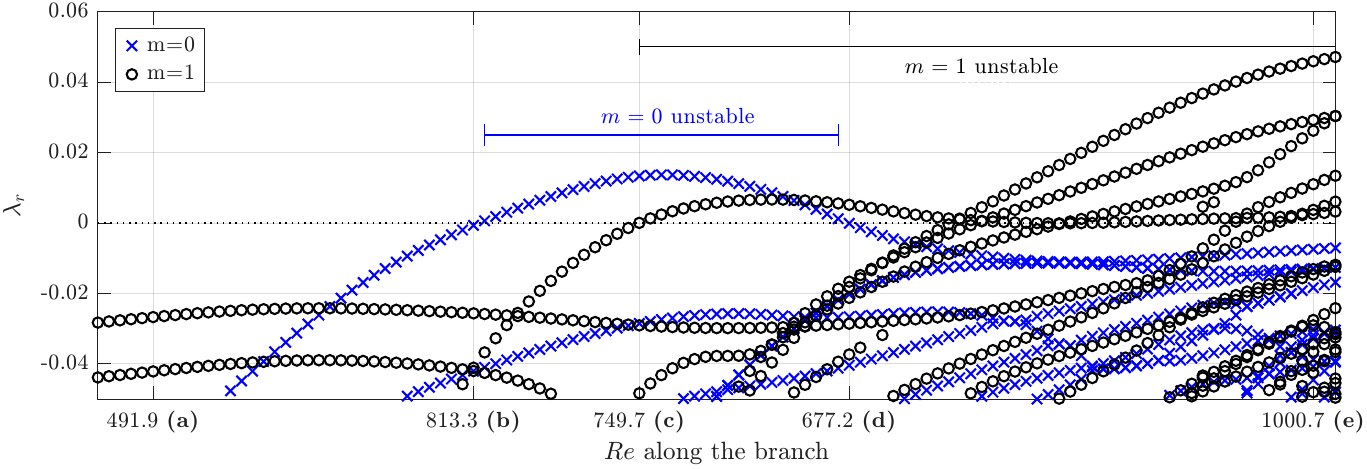}
\caption{Real part of the leading eigenvalue for $m=0$ and $m=1$ modes along the branch of steady solutions in figure \ref{fig:branchal0}. Mode $m=0$ in unstable only in the middle part of the branch connecting two saddle-node bifurcations at points (b) and (d). Mode $m=1$ is unstable from the point (c) onwards with growing number of unstable modes along the branch.}
\label{fig:stabalong}
\end{figure}

Contrary to \eqref{enpert}, the observable \eqref{eq:ez} is not null at the base flow and can be used to trace out the branch of steady solutions.
Two folds of the branch mark two saddle-node bifurcations in points (b) $Re=813.3$ and (d) $Re=677.2$ in figure \ref{fig:branchal0}. The base flows at the points marked in this figure are also shown in figure \ref{fig:slip} with the field for point (b) marking configuration where the recirculation bubble is born at the axis around $z\simeq 3$ (although the bubble does not have to appear exactly at the saddle-node point). Stability of the computed steady state solutions to azimuthal wavenumbers $m=0$ and $m=1$ is also presented in figure \ref{fig:branchal0}. Saddle-node points (b) and (d) correspond to the set of equations \eqref{nseq}-\eqref{contieq} being singular (when no arclength continuation is used). This is marked by a zero eigenvalue for $m=0$ present at these points (slight offset from the exactly zero value is caused by the fact that the continuation point is not exactly at the fold). The zero eigenvalue becomes unstable between the points (b) and (d) and the solution is stable to $m=0$ eigenmodes otherwise. This is marked by solid (stable) and dashed (unstable) lines in figure \ref{fig:branchal0}. The linear stability analysis for $m=1$ modes reveals that the identified branch becomes unstable around $Re=749.7$. From this point on, the branch is never stable to $m=1$ modes but the unstable mode switches from the one of $\lambda_i\approx-0.16$ to another one of higher angular frequency $\lambda_i\approx-0.33$. This switch of the modes is best seen in figure \ref{fig:stabalong} where $\lambda_r$ for both $m=0$ and $m=1$ is plotted along the steady branch together with $Re$ of interest from figure \ref{fig:slip} and \ref{fig:branchal0} marked. 

\begin{figure}
\centering
\includegraphics[width=0.8\textwidth]{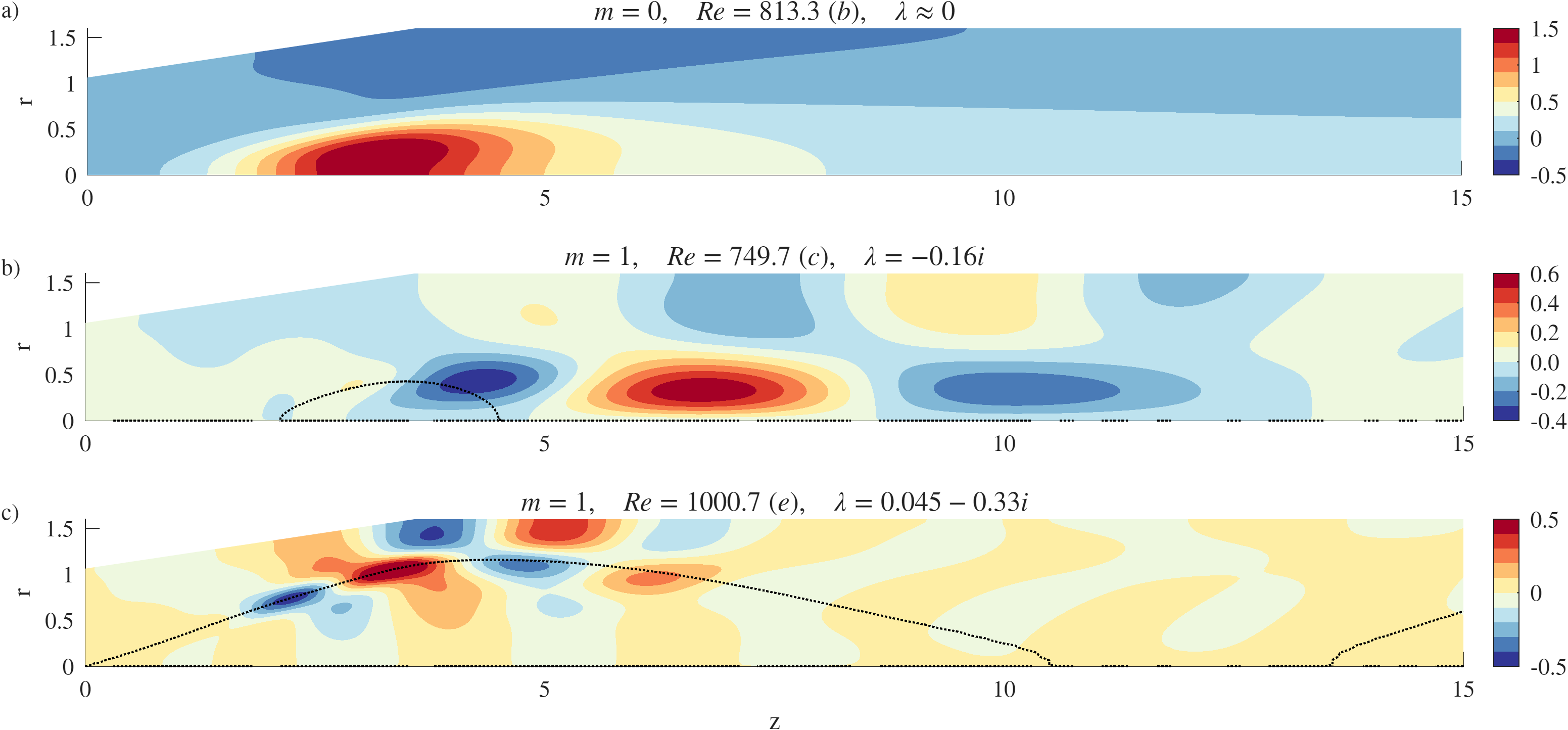}
\caption{Axial velocity of the selected eigenvectors at points chosen on the steady branch in figure \ref{fig:branchal0}: almost neutral eigenvector near the saddle-node point (a), the first helical mode to become unstable (b) and the unstable helical mode for the base flow characterised by a large recirculation bubble (c). The $\psi=0$ line marking the boundary of the recirculation bubble corresponding to figure \ref{fig:slip} is also marked with a dotted line.   }
\label{fig:evcslip}
\end{figure}

The axial velocities of the eigenvector for $Re=Re_{SN}=813.1$, $m=0$ (first fold), $Re=749.7$, $m=1$ (onset of $m=1$ instability) and $Re=1000.7$, $m=1$ (base flow characterised by a large recirculation bubble) are plotted in figure \ref{fig:evcslip}. The first eigenvector (figure \ref{fig:evcslip}a) corresponds to the incremental evolution of the axisymmetric steady state solution around the fold at $Re_{SN}$. Large amplitude at $z\approx4$ and $r<0.5$ is associated to the formation of the recirculation bubble which starts to be seen for the same $Re$ in figure \ref{fig:slip}b. The second eigenvector (figure \ref{fig:evcslip}b) is the first helical mode which becomes unstable at the branch of steady solutions. Its frequency and form does not seem to be easily related to the vortex rope dynamics. It can be however noted that the eigenvector seems to originate from the stagnation point formed at the head of the recirculation bubble and develops at the interface of the bubble. The recirculation bubble is marked in the figure with $\psi=0$ streamline extracted from figure \ref{fig:slip}c. The final eigenvector plotted (figure \ref{fig:evcslip}c) is again a helical mode, this time clearly developing at the boundary of the recirculation bubble. Its frequency $0.33\Omega$ is similar to the vortex rope frequency usually reported to be around $0.2-0.4\Omega$ \citep{amini2023upper}. Additionally, the fact that the vortex rope results from a helical instability mode of the recirculation bubble at the axis was postulated in the context of fully turbulent draft tube flow \citep{susan2009three}, and this observation provides an additional link to the present laminar computations.

\subsection*{Time integration}

Even though the helical mode is identified as an unstable mode of the steady solution, a question remains whether this steady state can be easily \agh{accessed} by time integration and whether this mode yields a saturated self-sustained state. In order to explore the flow dynamics that could potentially be reproduced experimentally a direct numerical simulations are conducted at a series of $Re$. By incrementing gradually $Re$, a new attracting region of phase space is identified for $Re>Re_{SN}$. 

\begin{figure}
    \centering
        \includegraphics[width=\linewidth]{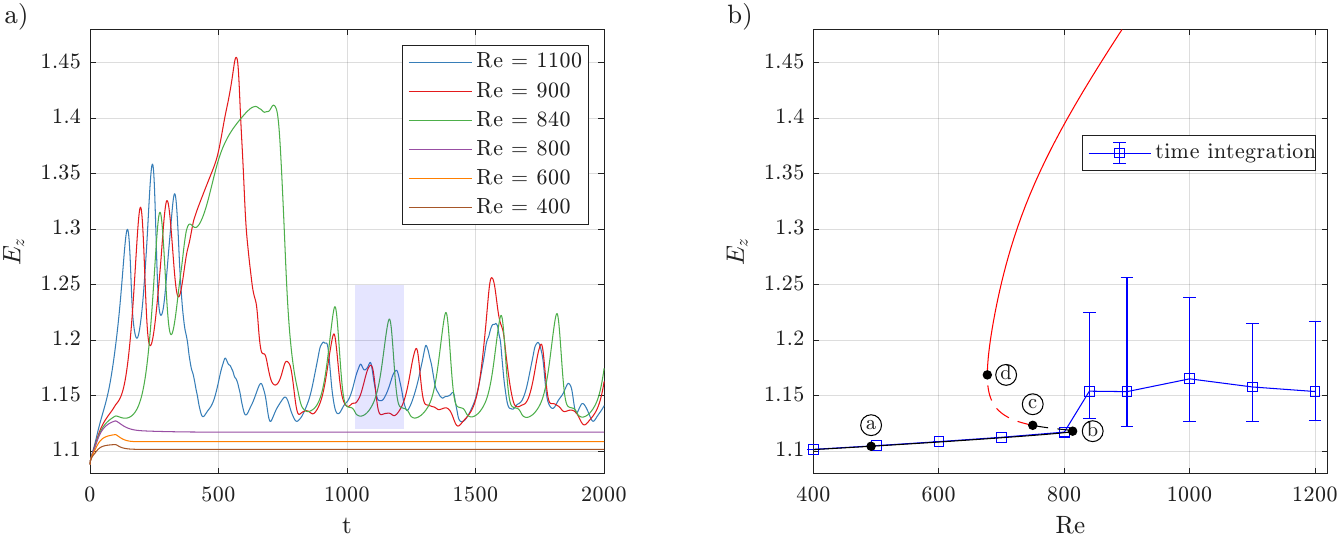}
    \caption{
    Time evolution of the observable \eqref{eq:ez} for a series of $Re$. Time series plotted in panel a) is summarised by its mean and extremum values (marked by a marker and an errorbar in panel b) for the second half of the signal ($t>1000$) and compared with the steady solution reported in figure \ref{fig:branchal0}. 
    Departure from the steady state solution is observed beyond the saddle-node bifurcation point ($Re>Re_{SN}$). Shaded region in panel a) corresponds to a series of snapshots of the solution presented in figure \ref{fig:breathing}.}
     \label{fig:ti}
\end{figure}

\begin{figure}
    \centering
            \raisebox{80pt}{a)}
    \includegraphics[width=0.8\linewidth]{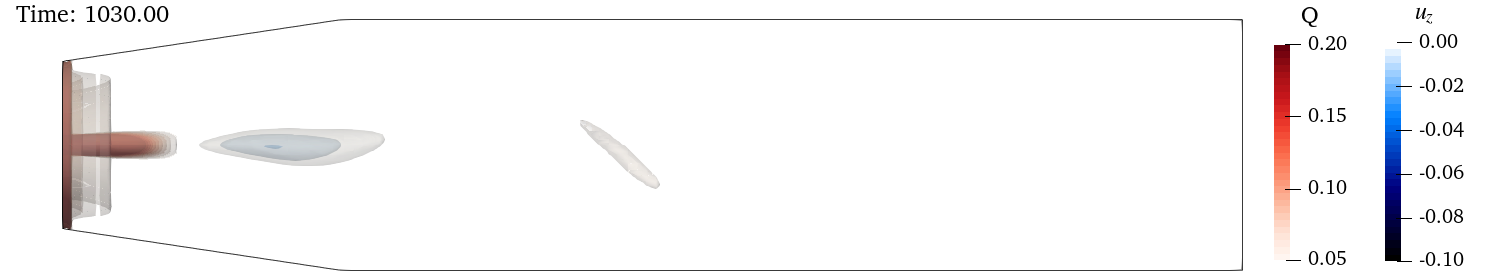}
    \newline
                \raisebox{80pt}{b)}
\includegraphics[width=0.8\linewidth]{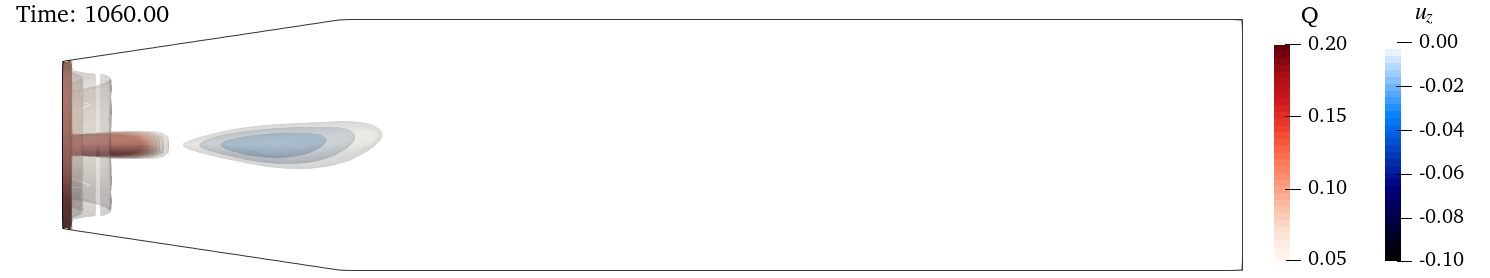}
\newline
\raisebox{80pt}{c)}
\includegraphics[width=0.8\linewidth]{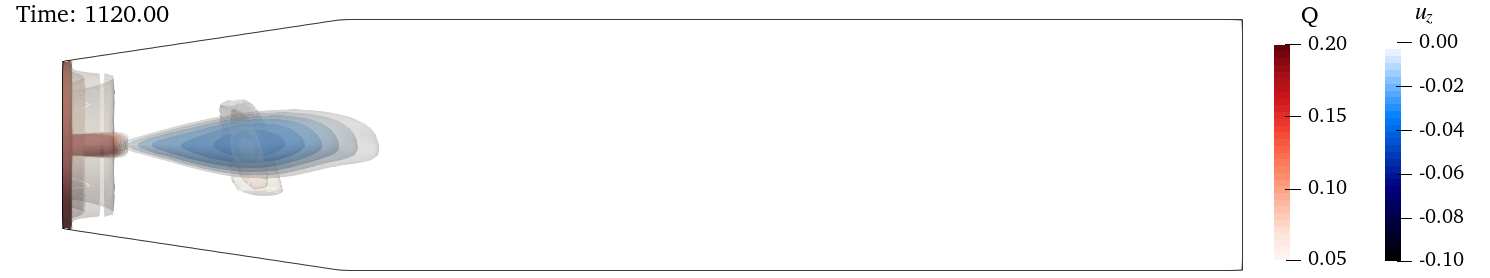}
\newline
\raisebox{80pt}{d)}
                    \includegraphics[width=0.8\linewidth]{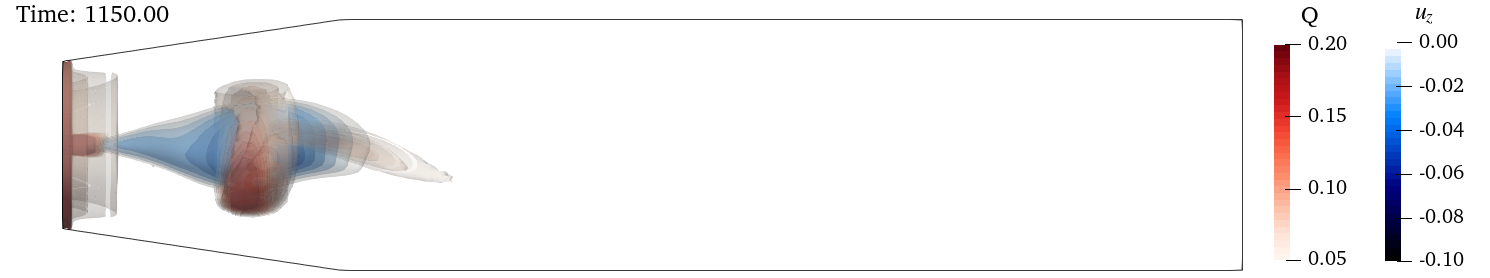}
                    \newline
                   \raisebox{80pt}{e)} 
                        \includegraphics[width=0.8\linewidth]{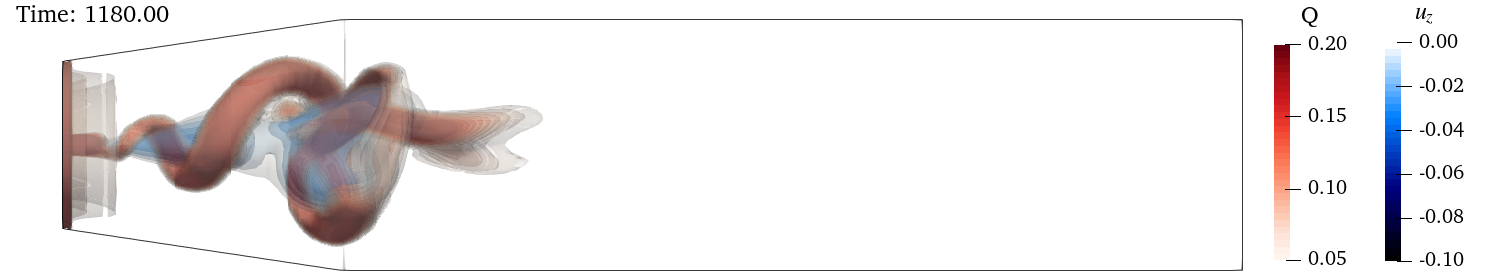}
                        \newline
                        \raisebox{80pt}{f)}
                            \includegraphics[width=0.8\linewidth]{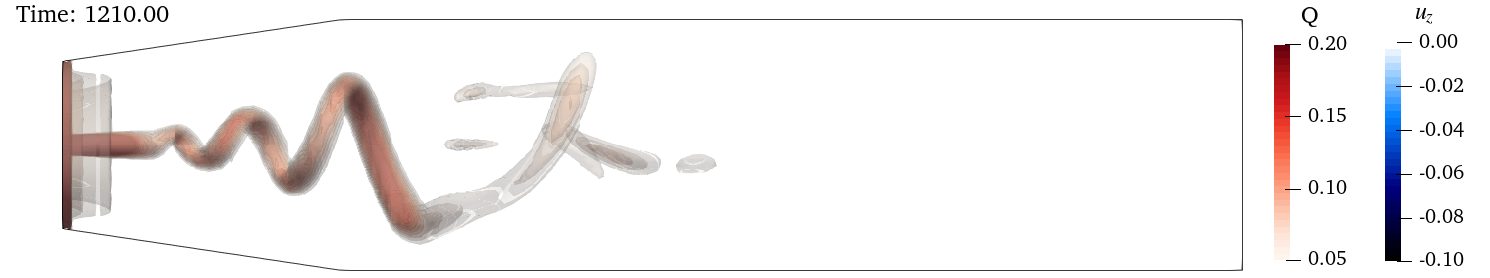}
                            \newline
                            \raisebox{80pt}{g)}
                                                        \includegraphics[width=0.8\linewidth]{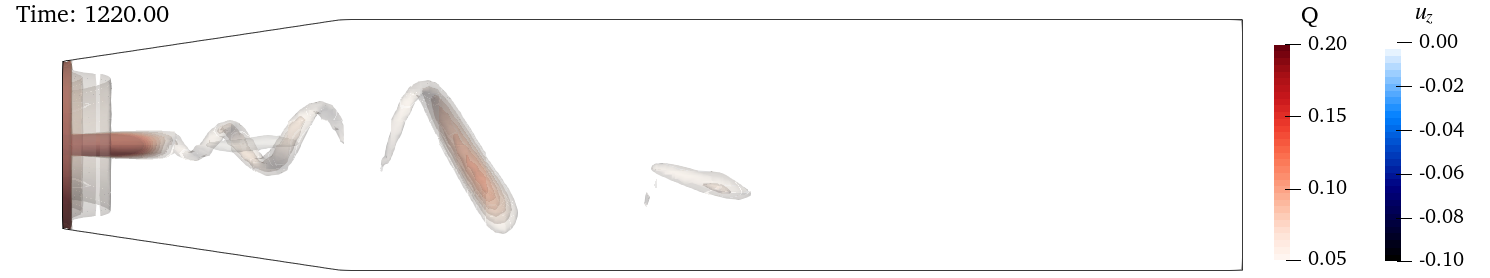}
                                                        \newline
    \caption{
    Snapshots of the unsteady solution at $Re=840$ visualised with isocontours of axial velocity in blue and Q-criterion in red. Panels a-d correspond to a gradual formation of the recirculation bubble and panels e-g to its bursting due to the formation of a helical mode around it. This process continues regularly as marked by oscillation of $E_z$ in figure \ref{fig:ti}. A video corresponding to the whole time integration signal plotted in figure \ref{fig:ti} at this $Re$ is provided as a supplementary material \cite{suppmat}. }
     \label{fig:breathing}
\end{figure}

Time series of the observable \eqref{eq:ez} is reported in figure \ref{fig:ti} for a series of $Re$. When $Re<Re_{SN}$, the time integration started from a zero uniform field converges to the steady state solution lying on the branch identified by arclength continuation. For $Re>Re_{SN}$, this branch is unstable to $m=1$, and the dynamics can visit the states in the vicinity of this branch at most transiently and will eventually diverge to reach another attracting state. Such a transient visit in the vicinity of the top branch unstable solutions takes place at $Re=840$ and $Re=900$ around $t=700$. It was checked (not shown here) that the instantaneous velocity field resembles closely figure \ref{fig:slip}e. The divergence from this state is along the unstable $m=1$ direction through a formation of a helical structure around the recirculation bubble. The formation of the $m=1$ mode destroys the recirculation bubble and an asymptotic state is attained for $t>1000$. It is marked by regular, although not exactly periodic, oscillations of $E_z$. These oscillations consist of subsequent formations and destructions of a recirculation bubble, although a smaller one compared to figure \ref{fig:slip}e. Snapshots of the instantaneous state are visualised using isocontours of axial velocity $u_z$ (negative values detect recirculation) and Q-criterion (positive values are used to visualise vortices \citep{Jeong_Hussain_1995}) in figure \ref{fig:breathing}. This regular formation and destruction of the recirculation bubble was reported before for a free Grabowski vortex in the direct numerical simulation of \cite{ruith2002,ruith2003three}. An additional video corresponding to figure \ref{fig:breathing} is provided as a supplementary material \cite{suppmat}. Mean and bounding values of $E_z(t>1000)$ are plotted in figure \ref{fig:ti}b. Good agreement for the mean $E_z$ value from time integration and arclength continuation is observed for $Re<Re_{SN}$. For $Re>Re_{SN}$ the flow exhibits a regular \textit{breathing} like dynamics centred around $E_z\approx1.15$. 

Further observations comparing the form of the vortex rope in no-slip and free slip simulations can be made. Changing the exterior wall boundary condition to free slip prevents the boundary layer from forming and leaves much more freedom for the vortex breakdown mode to develop. It is believed that this causes the rope to adopt a more conical shape, with a clearly defined opening angle visible in figure \ref{fig:breathing}f. This state is admittedly only a transient and the experimental vortex rope is of course a sustained state. Including the influence of turbulence on the large coherent structure of the rope, which is not taken into the account in these laminar computation, could be a way of sustaining the rope in the flow. It is also noted that the presumably spurious receptivity mode reported in figure \ref{fig:endti} is no longer present in the free slip simulations, as no boundary layer can form on a slip solid wall.

The fact of the existence of a saddle-node bifurcation and a sudden onset of large amplitude oscillatory dynamics past this point is reminiscent of the subcritical transition scenario often observed in shear flows. In this scenario, the large amplitude oscillatory solutions exist even below the critical point and can be accessed by a finite amplitude perturbation of the stable base flow. The coexistence of two stable locally attracting states gives rise to a hysteresis loop and the state reached by the dynamics depends on the initial condition of the simulation.

\subsubsection*{Hysteresis loop} 
%
%
%

\begin{figure}
    \centering
                                \includegraphics[width=\linewidth]{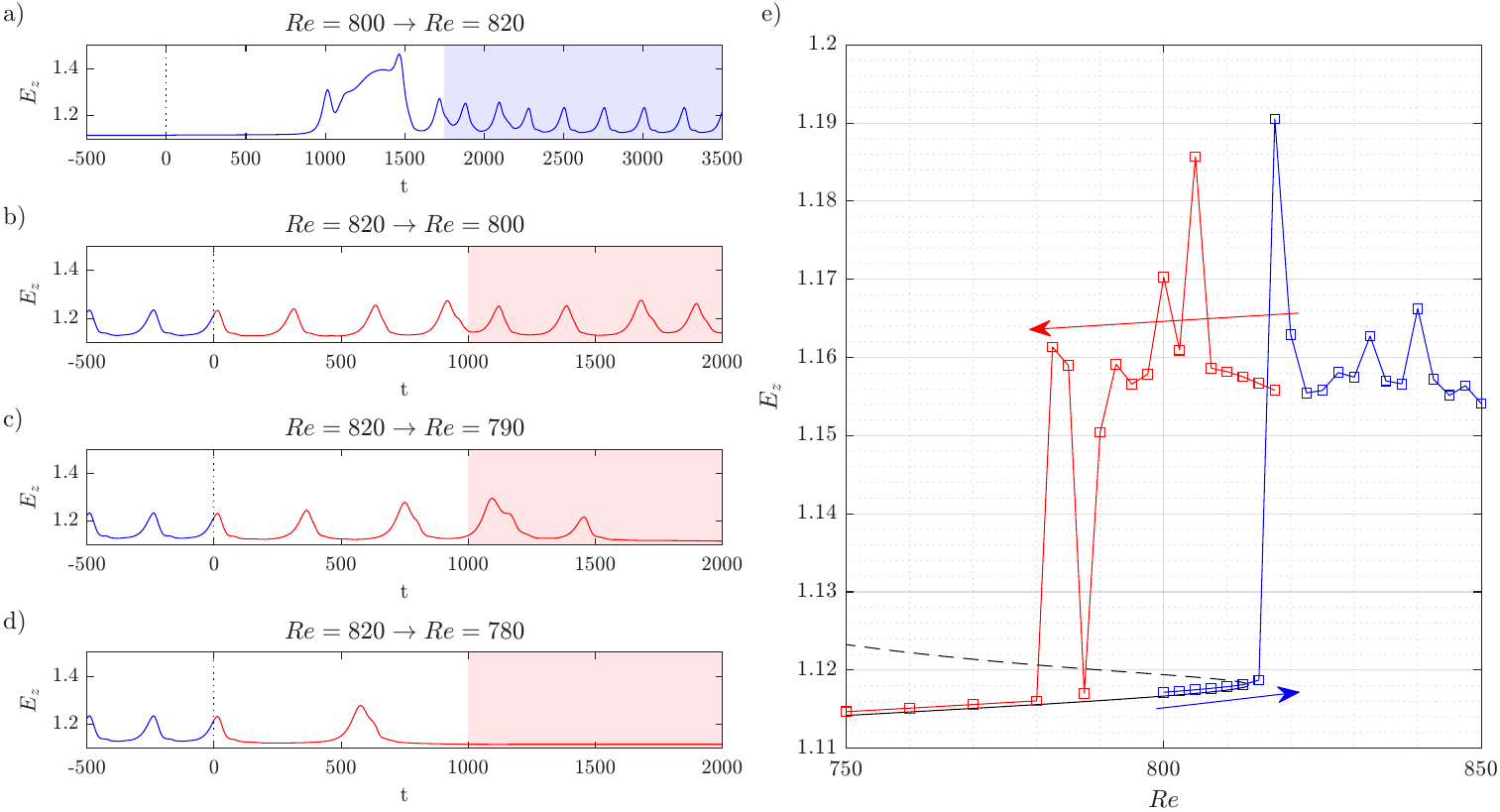}                 
    \caption{Time series of $E_z$ for simulations restarted from a converged steady solution at $Re=800$ (a) and from a snapshot 
    of the oscillatory state at $Re=820$ (b-d). The mean value, 
    averaged over the last 1750 time units for the increasing $Re$ case and over the last 1000 time units for the decreasing $Re$ case (shaded region in a-d) is shown in panel e) where a hysteresis loop is observed. Additional arrows show the direction of increasing (blue) or decreasing (red) $Re$ along the hysteresis loop. Stepwise change of $Re$ is performed at $t=0$ and is marked with black dotted line. }
     \label{fig:hyst}
\end{figure}

To test for the existence of such a hysteresis loop, we varied $Re$ both downward and upward, starting from different initial conditions. First, a stable steady state solution at $Re=800$ is used as the initial condition for a series of simulations for $Re>800$. Then, once a large amplitude oscillatory state is reached, $Re$ is decreased again in a series of simulations. If the hysteresis loop is easily reachable then, over at least some interval of $Re$, the dynamics should stay on the high energy level before collapsing back to the steady stable branch for some $Re<Re_{SN}$. This procedure and selected time series of $E_z$ are presented in figure \ref{fig:hyst}, with the time starting from $t=-500$, and $Re$ changed at $t=0$. 

Figure \ref{fig:hyst}a corresponds to a stepwise increase of $Re$ from $Re=800$ to $Re=820$. A long transient is followed by recovery of breathing dynamics described above. All simulations for the stepwise increase of $Re$ to a value $Re>Re_{SN}$ would follow a similar trend with a characteristic that the initial transient can be very long when $Re$ is close to the saddle-node point. The mean value of the second half of the signal ($t\in[1750,3500]$) is plotted in figure \ref{fig:hyst}e in blue for all of the simulations started from the steady state at $Re=800$. These values form a bottom branch of the hysteresis loop.  

The second branch of the hysteresis loop is found by conducting a series of simulations starting from the last snapshot of $Re=820$ simulation and then decreasing $Re$ in a stepwise manner. The state from which the simulations are restarted is similar to figure \ref{fig:breathing}d. It corresponds to a recirculation bubble which has not yet grown to its final size before bursting. This is also marked by the value of $E_z$ just before a local minimum at the end of the time integration reported in figure \ref{fig:hyst}a. Decreasing $Re$ from $Re=820$ to $Re=800$ (figure \ref{fig:hyst}b) retains the breathing dynamics, whereas a similar decrease to $Re=790$ (figure \ref{fig:hyst}c) results in a few bubble burst before recovering the steady bottom branch solution. This scenario is even more pronounced when the decrease to $Re=780$ (figure \ref{fig:hyst}d) is made and only two bursts are followed by recovery of the steady solution.  The mean value of the second half of the signal ($t\in[1000,2000]$) is plotted in figure \ref{fig:hyst}e in red. These values form a top branch of the hysteresis loop. The exact value of $Re$ for the tipping point of the top branch (where the steady solution is recovered after a long transient) naturally depends on the initial condition and on the length of the time integration. The mechanism of relaminarisation could follow the edge catastrophe scenario, as the top branch attractor expands in phase space, it collides with the edge state, becomes leaky, and trajectories can escape toward the laminar steady branch. A similar mechanism has been described in the case of pipe flow by \cite{avila2013streamwise}. This direction is not pursued in this work, instead, we limit our discussion to the general observation on the existence of the hysteresis loop in the interval of around $Re\in(800,813.3)$. A certain refinement of the mean values plotted in figure \ref{fig:hyst}e could be obtained by performing longer time integrations (e.g., two or four times longer), but the qualitative observation on the existence of the hysteresis would remain unchanged.

\subsubsection*{Ghost solutions}

\begin{figure}
    \centering
         \includegraphics[width=\linewidth]{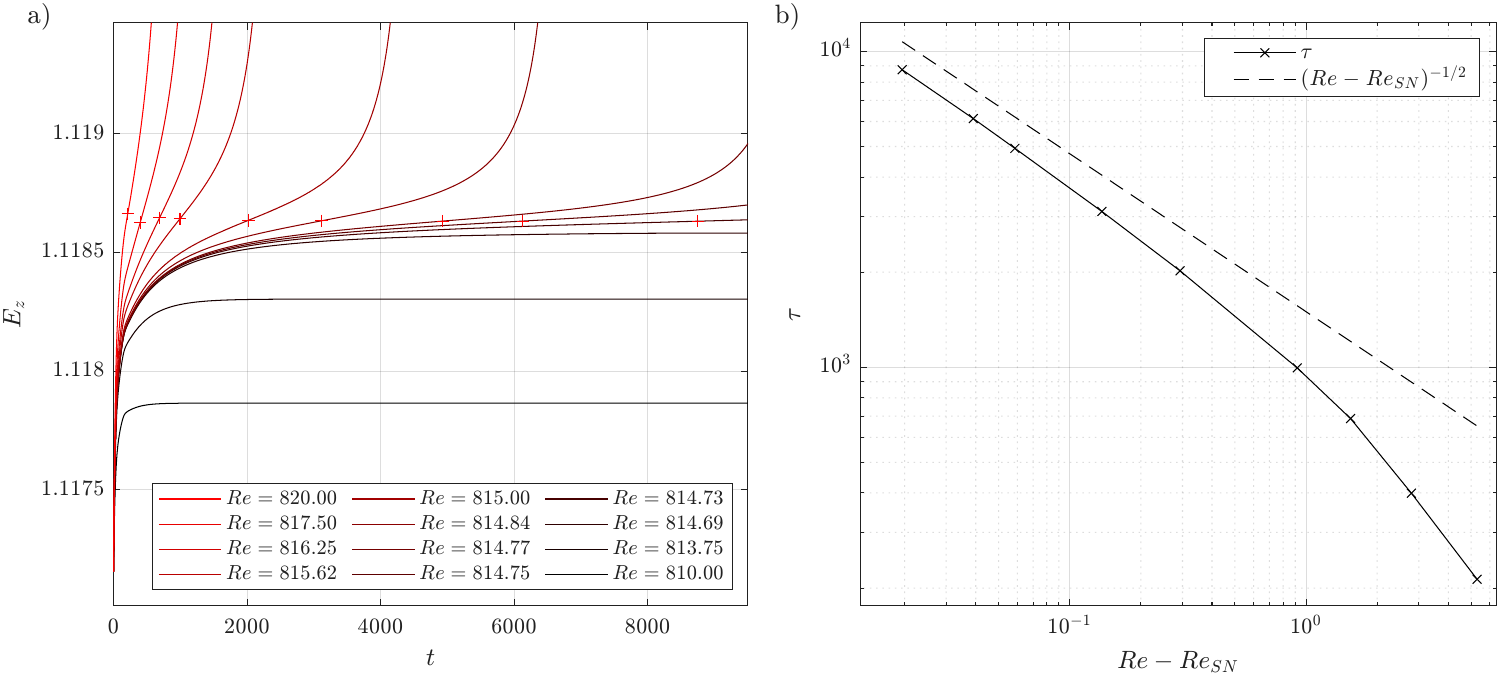}
    \caption{Time series of $E_z$ very close to the saddle-node bifurcation point a) and the time spent near the ghost solution b) measured as the time from the start of the time integration until the inflection point of $E_z(t)$ marked in in a red cross in panel a). Good agreement of the time spent in the vicinity of the ghost solution with theoretically predicted scaling (dashed line) is observed. 
}
     \label{fig:ghost}
\end{figure}

As discussed in the previous section, when a time integration is performed very close to the saddle-node point, the trajectory can spend a considerable amount of time in the region of phase space associated with the saddle-node bifurcation. This phenomenon is sometimes referred to as an existence of a \textit{ghost solution} or a \textit{ghost} of the solutions annihilated in the bifurcation \citep{strogatz2001nonlinear}. This ghost solution, while it is not an exact steady solution of the governing equations, it is almost a solution and therefore influences the nearby dynamics. Numerous examples of fluid and structural systems where the ghost solutions are present are given by \cite{zheng2025ghost}. An analysis of the normal form on the saddle-node bifurcation provides an exact solution for the evolution near the bifurcation point with $A\propto\tan(t)$ for some observable $A$ and the time spent in the vicinity of the ghost $\tau\propto{1}/{\sqrt{\mu}}$ for a control parameter $\mu$. Both scalings, noted also by \cite{zheng2025ghost}, can be appreciated in figure \ref{fig:ghost}. Here, the proximity to the saddle-node point is $Re-Re_{SN}$. The value $Re_{SN}$ is estimated with a bisection procedure and  taken as a mean of the last two bisection steps $Re_{SN,DNS}=(814.69+814.73)/2=814.71$. The time spent near the ghost solution is measured as the time required for $E_z(t)$ to reach an inflection point identified by a minimum of {${dE_z}/{dt}$} (not shown) saved during the simulation. This inflection point is marked with a red cross in figure \ref{fig:ghost}a for $Re>Re_{SN,DNS}$. For $Re<Re_{SN,DNS}$, there is no inflection point since $E_z$ converges to the value corresponding to the steady state solution. The fact that $Re_{SN,DNS}$ is slightly different from the value of $813.3$ found with Newton method in FreeFem++ is attributed to the finite size of the mesh and a finite time step used in Nek5000. The difference is however negligibly small.  Presence of the ghost solution explains a long plateau of $E_z$ at $Re=820$ for $t<800$ in figure \ref{fig:hyst}a.

\section{Higher flow rate operating point} \label{sec:dop}
\begin{figure}
    \centering
        \includegraphics[width=\linewidth]{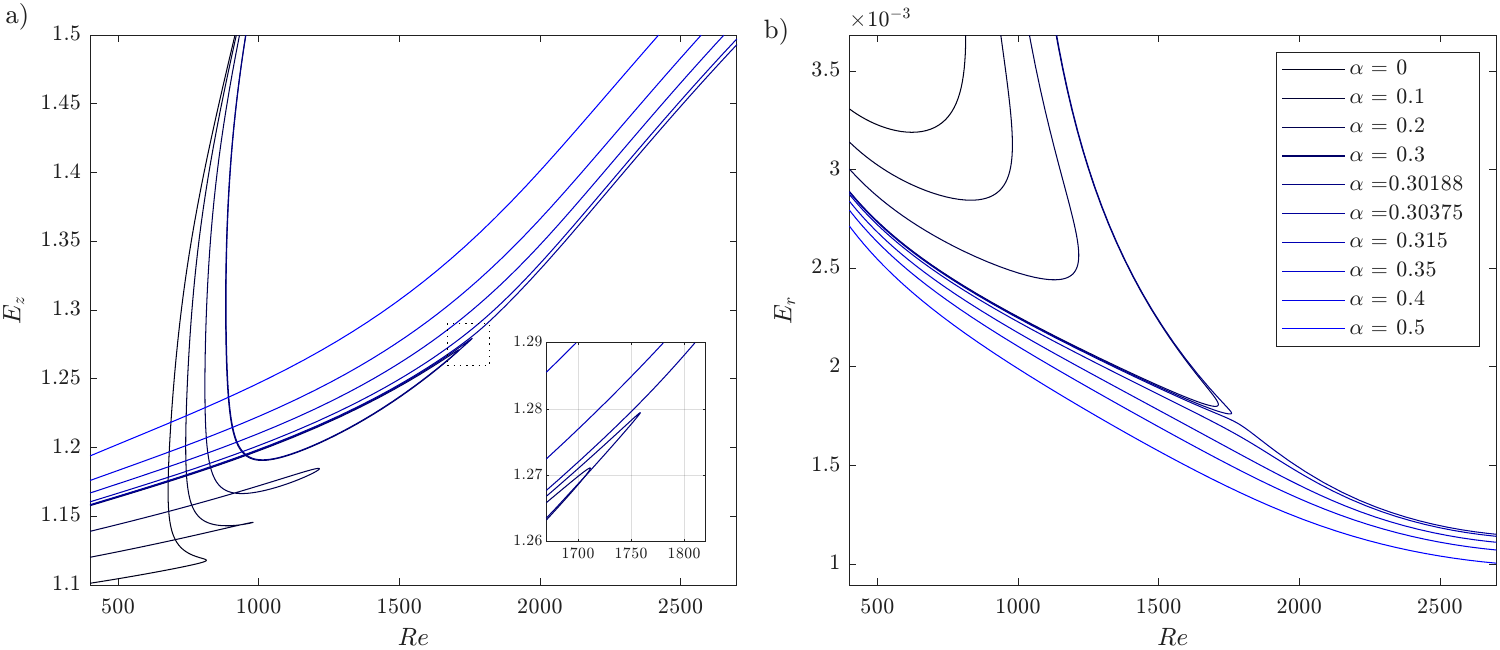}
    \caption{Branch of steady solution found for a series of interpolation parameters $\alpha$ in eq. \eqref{eq:interp}. In panel a) the observable \eqref{eq:ez} is plotted but a clearer picture can be  obtained with  observable \eqref{eq:er} in (b). Inset in panel a) is a zoom on the region bordered by the dotted line in the vicinity of the cusp forming on the branch. Legend in panel b) is common to both panels.
     }
     \label{fig:ezer}
\end{figure}

Computations for the free slip wall are repeated \revision{\agh{in this section for a series of inflow conditions between 0.92 BEP and 0.98BEP. A continuous linear interpolation between these two profiles is performed with a parameter $\alpha$, as in section}} \ref{sec:baseflow}, equation \eqref{eq:interp}.
The branch of steady solutions as in figure \ref{fig:branchal0} is found for a range of $\alpha$. Evolution of this branch with $\alpha$ is shown in figure \ref{fig:ezer}. The steady solution is always computed first at $Re=400$ and the arclength continuation procedure is started from this point for all $\alpha$. For growing $\alpha$ the branch traced by $E_z$ starts to intersect itself (figure \ref{fig:ezer}a for $\alpha = 0.1, 0.2, 0.3$ and 0.30188). This is however only due to the choice of observable and the same value of $E_z$ does not mean that the whole velocity field is the same. This is seen in figure \ref{fig:ezer}b where the branch is traced using an alternative observable,
\begin{equation} \label{eq:er}
E_{r}=\frac{1}{2}\int_0^{15}\int_0^{2\pi}\int_0^{R(z)}u_r^2\; r\,\mathrm dr\,\mathrm d\theta\,\mathrm dz,
\end{equation}
and no self-intersection is observed. Independent on the choice of observable, the branch undergoes a sudden change in direction around $\alpha=0.3$. A precise $\alpha$ value is determined with a bisection procedure and is determined as an average of two last bisection steps $\alpha_c=(0.30188+0.30375)/2=0.3028$. At this point, the operating point would correspond to 0.94BEP. With $\alpha$ growing from 0 to $\alpha_c$, the turn of the steady branch becomes increasingly sharp and then, unexpectedly, the fold disappears with the branch displaying no fold at all for \revision{$\alpha>\alpha_c$ (for $\alpha>0.5$ the branch displays no fold but this parameter range was omitted from the representation in figure \ref{fig:ezer} in order to focus on the values in the vicinity of $\alpha_c$).}   
This behaviour is reminiscent of an imperfect transcritical bifurcation with an imperfection 
parameter $\alpha$ \citep{charru2011hydrodynamic}. Branch unfolding through an imperfect bifurcation was previously reported in a swirling jet by \cite{lopez1994bifurcation}. In that case, a pitchfork bifurcation takes place at finite $Re$ with the swirl intensity acting as an imperfection parameter. The case of a transcritical bifurcation was also studied \agh{by }\cite{wangPof1996a,elenavphd} \agh{and }\cite{wang1997effect}\revision{\agh{ showed that a slight viscosity breaks this inviscid transcritical bifurcation into two folded branches separated by a finite gap.}} However, those works \agh{focus on the bifurcations in the inviscid limit}. Present results \revision{document the swirling jet transcritical bifurcation at finite $Re$, similarly to findings of \cite{QiaoPof2025} in Lamb-Oseen vortex. \agh{Contrary to the results of }\cite{QiaoPof2025}\agh{, who reports a series of transcritical bifurcations for a range of swirl numbers and two $Re$ values, the present transcritical bifurcation is structurally unstable - it takes place only at a specific combination of swirl (effectively turbine operating point) and $Re$. Departure from this critical swirl causes the bifurcation to decay into two disjoint solution branches, as discussed below.}}

\begin{figure}
    \centering
    \includegraphics[width=\linewidth]{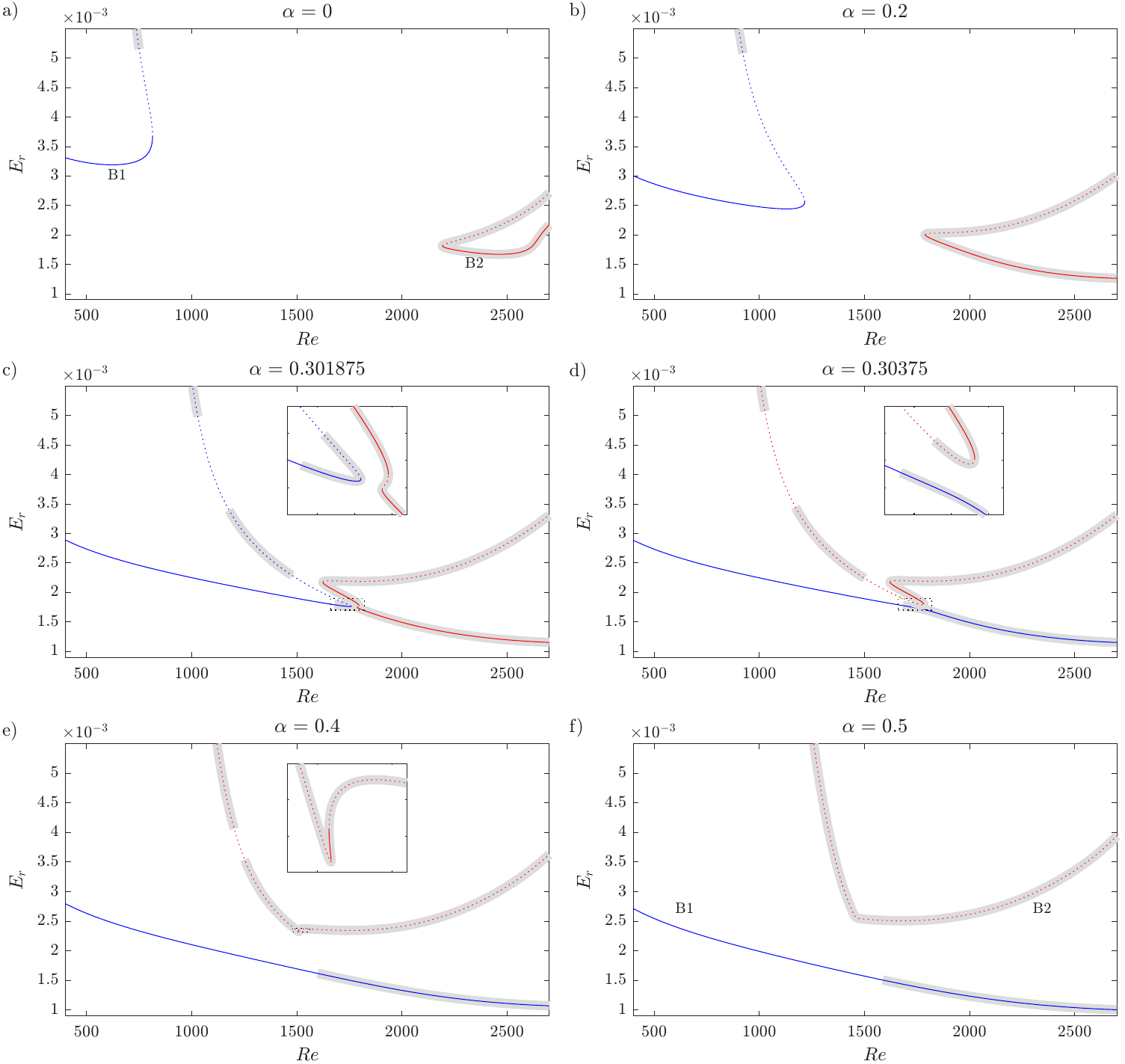}
    \caption{Two branches of steady solutions identified with arclength continuation procedure for a range of interpolation parameters $\alpha$. Collision of the branches between points $\alpha=0.301875$ and $\alpha=0.30375$ (panels c-d) marks a transcritical bifurcation. Insets in the corresponding panels correspond to a region bordered by a black dotted line magnified for better clarity. Stability to $m=0$ modes is marked with solid (stable) and dotted (unstable) lines. Instability to $m=1$ modes in marked with a shaded region. }
        \label{fig:6panel}
\end{figure}

A transcritical bifurcation postulated above would imply that there exists another branch of steady state solutions and the solutions shown in figure \ref{fig:ezer} collide with this branch in a transcritical bifurcation. In order to identify these additional branches, the solution from the original branch at $\alpha=0.4>\alpha_c$, $Re=2700$ is used as an initial guess for the arclength continuation, with $\alpha$ decreased to a value $\alpha<\alpha_c$. The result of the continuation procedure is depicted in figure \ref{fig:6panel}a-c. The original branch presented in figure \ref{fig:ezer} is now plotted in blue and is labelled as B1 and a newly found branch is plotted in red and labelled as B2. Instability to $m=0$ modes is also indicated with a dotted line and stability is observed to change at the fold points, as expected. Instability to $m=1$ modes is marked with a shaded region in the same figure. It is computed by finding 40 eigenvalues closest to the complex shift with a series of shifts (-2.1i,-1.7i,...,1.9i) at each arclength continuation point. Generally, $m=1$ is always unstable for sufficiently large $Re$, and the second branch (B2) is more unstable to $m=1$ than B1. The stability can alternate with instability along the branches. 
Solutions on branch B2 for $\alpha>\alpha_c$ are computed similarly, but using as initial guess the solutions for $\alpha=0.3$, $Re=1700$ with high $E_z$ and $E_r$ (point not visible in figure \ref{fig:ezer}).
 The result of this computation is plotted in figure \ref{fig:6panel}d-f.

\begin{figure}
    \centering
    \includegraphics[width=0.45\linewidth]{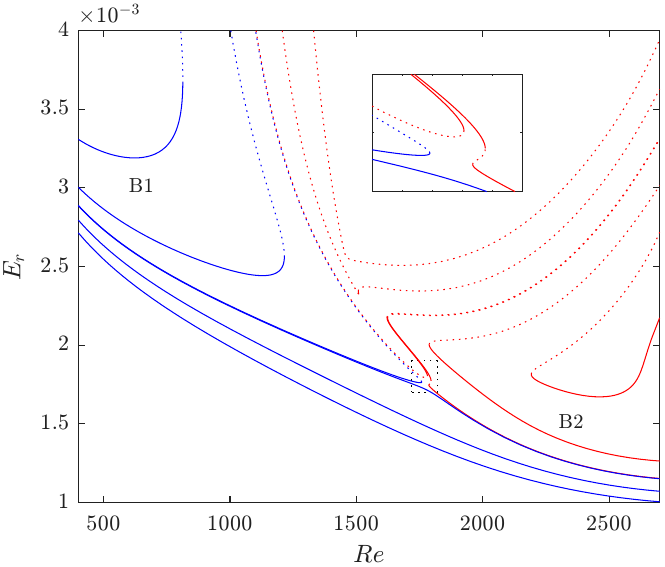}
\caption{All branches reported in figure \ref{fig:6panel} plotted together. Region where the transcritical bifurcation takes place is magnified in the inset. Stability to $m=0$ modes is marked as in previous figure. }
    \label{fig:alltog}
\end{figure}

For growing $\alpha$, branches B1 and B2 approach each other. The collision with branch B1 is initiated by the formation of two additional folds on branch B2, as seen in figure \ref{fig:6panel}c and its inset. Similar scenario for a transcritical bifurcation preceded by formation of a double fold was described in Taylor-Couette flow by \cite{benjamin1978bifurcation}. At $\alpha_c$, steady solutions collide, and B1 inherits part of B2 for $\alpha>\alpha_c$. The collision of these two solutions branches and subsequent exchange of stability occurs via a transcritical bifurcation.
 Two folds that form on branch B2 in figure \ref{fig:6panel}c persist for $\alpha>\alpha_c$, but they are themselves unfolded 
 for $\alpha=0.5$ as seen in figure \ref{fig:6panel}f. A detailed explanation of this unfolding is left for a separate future study.
 All branches reported in figure \ref{fig:6panel} are then plotted together in figure \ref{fig:alltog} where the transcritical bifurcation point can be seen at $Re\approx1775$ and $E_r=1.76\times10^{-3}$.  

The scenario of branch unfolding via a transcritical bifurcation compares favourably with the experimental evidence of the vortex rope. At lower BEP ($\alpha=0$, high swirl), a high amplitude vortex rope emerges suddenly when a control parameter threshold is crossed   
(in the case of present laminar computation, it is $Re_{SN}$). This scenario corresponds to the most dangerous subcritical configuration. In our laminar computations, the vortex rope takes form of regular breathing dynamics, consisting of the repeated formation and destruction of the recirculation bubble at the axis and the vortex rope at its boundary. On the other hand, high BEP conditions (high $\alpha$, less swirl) correspond to a much safer situation where the branch of steady solutions does not fold and can be continued to large values of the control parameter. This is also the case in experiment where flow regimes close to the 1.0BEP are at the turbine design point and should be void of large amplitude oscillations.


\section{Conclusion and outlooks} \label{sec:summ}
In this work, an industrial flow inside the draft tube of a Francis turbine was studied. The draft tube geometry is simplified into diffuser-pipe elements following the axisymmetric domain proposed by \cite{susan2010analysis} with an inflow profile based on the experimental measurements reported therein. A simplification relative to real, fully turbulent flows is adopted here, as the results presented in this work focus on laminar flow. Nevertheless, several key characteristics of vortex breakdown that are consistent with experimental observations are recovered and are summarised below.

For a no-slip wall boundary conditions, an azimuthal wavenumber $m=1$ mode was found to be caused by a supercritical Hopf bifurcation at $Re_c$. With increasing $Re$, the saturation amplitude of the mode follows a familiar square root scaling. Whenever $Re>Re_c$, the dynamics on the saturated branch of oscillatory solutions is dominated by the Hopf bifurcation frequency, and no other frequencies (apart from harmonics of the dominant frequency) are detected in the time signal, even when the base flow is linearly unstable to multiple modes with different frequencies. The scenario of the formation of the vortex rope is therefore simple and very robust. This is a characteristic shared with an experimental vortex rope which is a robust $m=1$ self-sustained structure of one dominant frequency. What was found to be far from the experimental evidence was the cylindrical shape of the $m=1$ mode and its high frequency. Even though these discrepancies are likely linked to the laminar nature of the computations, a way to improve the results by changing the wall boundary condition from no-slip to free slip was proposed in this work.

For the free slip boundary condition, the branch of steady solutions identified at low $Re$ was found to undergo two folds, corresponding to saddle-node bifurcations. Time integration for $Re$ beyond the saddle-node bifurcation at $Re_{SN}$ has shown that the steady branch is not an attractor any more. Beyond the first fold, the branch is unstable to $m=0$, $m=1$ or both modes simultaneously. Rather than converging to the steady solution, the dynamics is characterised by regular, although not strictly periodic, events of recirculation bubble growth and its subsequent bursting due to a helical mode formed at its periphery. This \textit{breathing} dynamics was found to be sustained for all $Re>Re_{SN}$ and also for $Re<Re_{SN}$ provided that a suitable initial condition is chosen. A small hysteresis loop was identified in this way. 

The results obtained with the free slip wall condition were found to be more consistent with the experimental evidence of the vortex rope. Its shape is following a cone with a clear opening angle and the modes which the steady state is unstable to, are characterised by angular frequencies $0.1-0.4\Omega$, in line with experimental evidence. These improvements are likely due to the absence of a wall boundary layer in the free slip case, which allows the mode greater freedom to develop its spatial structure. A property lost by modifying the boundary condition was that the rope is no longer a self-sustained state but is only a regularly visited transient solution. 

A desirable outcome would be a simulated vortex rope that combines the key features of both no-slip and free slip results reported here: i) a self-sustained $m=1$ mode, ii) a conical shape and iii) a frequency around $0.3\Omega$. A natural step forward would be to include the turbulence interaction in the modelling. If the computationally expensive large eddy simulation or direct numerical simulation are to be avoided, a possible alternative is be to adopt an eddy viscosity approach and test its suitability in capturing the unsteady vortex rope.
  Initial results that identify the vortex rope as a unstable mode of a turbulent flow are already available \citep{pasche2017part,head_paper_1}. A more in-depth understanding of the bifurcation scenario and a parametric study are still lacking.  

Exploring the space of different turbine operating points was also investigated. Increasing the turbine operating point from 0.92BEP to 0.98BEP (effectively decreasing the inflow swirl) removes one of the folds of the steady state branch via a transcritical bifurcation. \revision{This type of bifurcation was reported for a swirling jet by \cite{wangPof1996a} in the inviscid regime and by \cite{QiaoPof2025} at finite $Re$. Factors favouring this vortex breakdown scenario could be connected to the fact, that the flow considered here is atypical, being characterised by the divergent geometry section and a particularly rich inflow profile.} Outlooks include a weakly nonlinear analysis \cite{Meliga_Gallaire_Chomaz_2012} and constructing a self-consistent model \citep{mantivc2014self,Yim19SelfConsitent} for the bifurcations identified. 
A sensitivity study on the shape of the domain and identification of the invariant solutions responsible for breathing dynamics are additional possible extensions of present work. \revision{A systematic exploration of the two dimensional parameter space formed by turbine operating point and the Reynolds number for both free slip and no-slip wall computations \agh{and an analysis based on Reynolds-Orr equations} (e.g. \cite{Qiao_Shi_Meng_Wang_Liu_2025}) \agh{would map how the stability and bifurcation scenarios depend on the turbine operating regime. }}

\section*{Acknowledgments}

The authors would like to thank François Gallaire and Edouard Boujo for fruitful discussions.


\section*{Appendix A : Mesh resolution sensitivity}

\begin{figure}
\includegraphics[width=\textwidth]{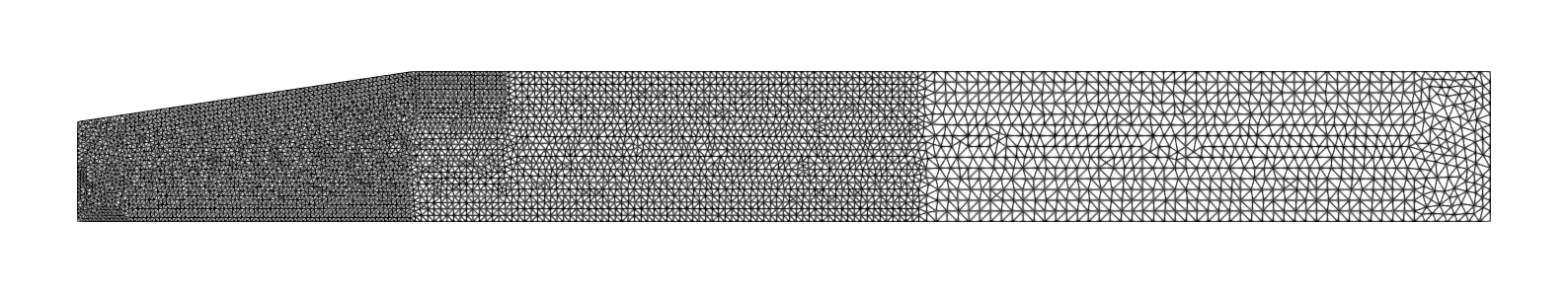}
\includegraphics[width=\textwidth]{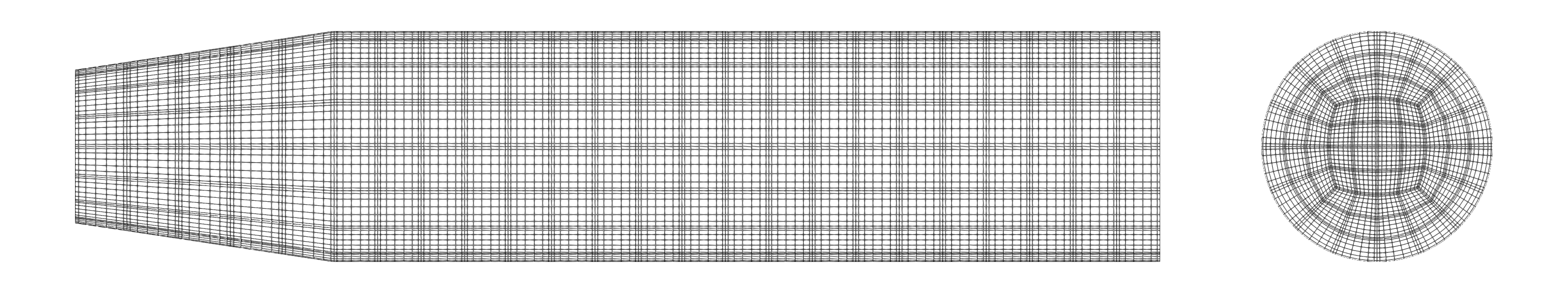}
\caption{FreeFem++ mesh consisting of $10\;281$ triangles (top) and Nek5000 mesh consisting of 1536 hexahedral elements (bottom). These are the coarsest meshes used for the respective codes. Meshes are refined uniformly when generating their finer counterparts. }
\label{fig:meshes}
\end{figure}

\begin{figure}
\includegraphics[width=\textwidth]{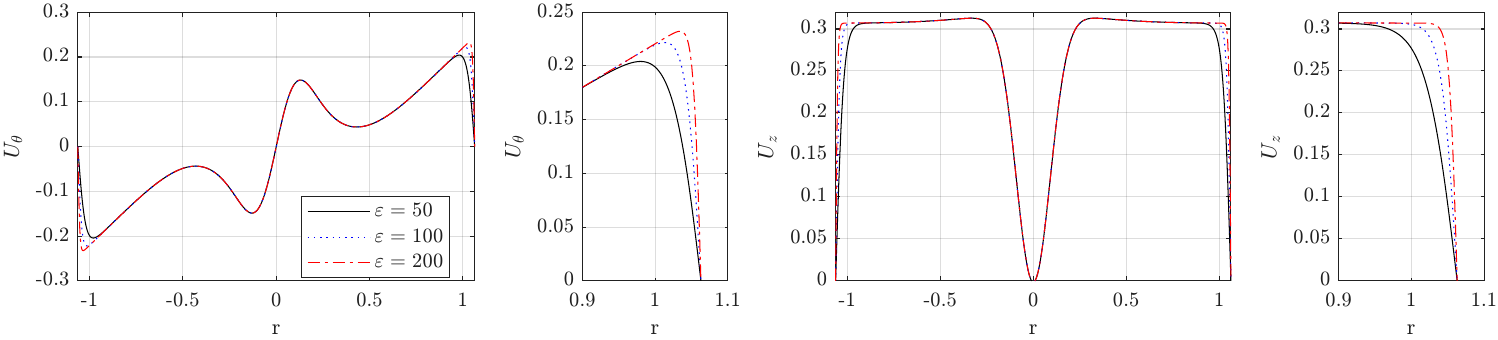}
\caption{\revision{Comparison of the width of regularisation controlled by parameter $\varepsilon$ for the 0.92 BEP inflow. The regularisation function and inflow profile are defined in section \ref{sec2}.} }
\label{fig:widtheps}
\end{figure}

Table \ref{tab:mc} presents sensitivity of the critical thresholds in $Re$ to the mesh resolution used in FreeFem++ and to the regularisation width of the inflow profile. The coarsest FreeFem++ mesh is also shown in figure \ref{fig:meshes}. We find $Re_c\approx2300$ and $Re_{SN}\approx800$ independent of the mesh used, however, the regularisation width influences the $Re$ threshold. The mesh with $n_t=23\;594$, $\varepsilon=50$ is used throughout the article.  For the computations on two finest meshes the interface to PETSc/SLEPc as available in FreeFem++ is used. \revision{The width of the regularisation is additionally presented in figure \ref{fig:widtheps} for the $\varepsilon$ values listed in table \ref{tab:mc} for the inlet profile corresponding to 0.92 BEP (cf. figure \ref{inlet}).  }

 The coarsest hexahedral mesh of 1536 elements is also shown in figure \ref{fig:meshes}. Two additional uniformly refined meshes with respect to this one, 6240 and 21472 elements, are also used in the present work as referred to in the text. 

%
%

\begin{table}[H]
\centering
\begin{tabular}{c|c|c|c|c|c|cccc}
{Reg. width ($\varepsilon$)} & \multicolumn{5}{c|}{50} & 100 & 200\\ \hline
{No. triangles ($n_t$)} & $10\,281$& $ 23\,594$ & $ 54\,181$  & $131\,115$ & $275\,516$  & \multicolumn{2}{c}{$23\,594$}\\
\hline \hline
no-slip $Re_c$ &2388.21& 2340.86 & 2333.14 & 2331.38 & 2331.19&\ 2149.63 &2086.21\\
free slip $Re_{SN}$ &813.24& \agh{813.3} &812.87&812.82&812.91&709.75 & 673.2 \\
\end{tabular}
\caption{Mesh convergence of the critical $Re$ for the Hopf bifurcation in the no-slip case, and the first fold of the steady branch in the free slip case, for different total numbers of triangles $n_t$ at fixed regularisation  width $\varepsilon=50$ (first three columns) and for fixed mesh size $n_t=23\;594$ with varying regularisation width (last two columns, larger $\varepsilon$ corresponds to a steeper decay). }
\label{tab:mc}
\end{table}

\section*{Appendix B : Method of Lagrange multipliers}
Free slip boundary condition consist of requiring that the velocity field $(u_r,u_{\theta},u_z)$ in the computational domain $D$ 
 with boundary $\Gamma=\partial D$, satisfies a zero-shear stress condition on the portion $\Gamma_{slip} \subset \Gamma$ where the condition is applied as
\begin{equation} \label{eq:nostrt}
(\bm \tau \cdot \bm n) \cdot \bm t = 0,
\end{equation}
and there is no flow through the border 
\begin{equation} \label{eq:noperm}
\bm u \cdot \bm n =0, 
\end{equation}
with the stress tensor $\bm \tau$, unit normal $\bm n$ and tangent vectors $\bm t$ to the boundary. Since the border $\Gamma_{slip}$ consists of straight line segments in the $(r,z)$ plane, the condition \eqref{eq:nostrt} gives for the in-plane velocity, to
\begin{equation} \label{eq:nostrtco}
\frac{\partial}{\partial \bm n} (\bm u \cdot \bm t)=0,
\end{equation}
i.e. there is no gradient of inplane velocity in the wall normal direction. The free slip condition on the azimuthal component $(\bm\tau\cdot\bm n)\cdot\bm e_\theta=0$, yields,
\begin{equation} \label{eq:nostrth}
n_r\!\left(\frac{\partial u_{\theta}}{\partial r}-\frac{u_{\theta}}{r}\right)+
n_z\frac{\partial u_{\theta}}{\partial z}=0.
\end{equation}

Condition \eqref{eq:noperm} simplifies 
\begin{equation}
[u_r,\;u_z]\cdot[n_r,\;n_z]=0.
\end{equation}
A natural condition imposed by the finite element {weak} formulation {for the diffusion operator of $u_\theta$} is 
$\left[ {\partial u_{\theta}}/{\partial r},\;{\partial u_{\theta}}/{\partial z}\right]\cdot[n_r,\;n_z]=0$, {i.e. $\partial_n u_\theta=0$.} Hence, an additional integral should be added to the weak formulation
\begin{equation}
\int_{\Gamma_{slip}}\nu \frac{u_{\theta}}{r}\,n_r\,v_{\theta}\,r {\mathrm{d}}s,
\end{equation}
where $\nu$ is the kinematic viscosity and $v_{\theta}$ is a test function, 
{in order}
 to satisfy no shear stress in tangential direction \eqref{eq:nostrth}. Finite element discretisation also naturally satisfies \eqref{eq:nostrtco} if \eqref{eq:noperm} is enforced. The latter is done by introducing a Lagrange multiplier field $l$ with a test function $w$, both discretised on P2 triangular elements. An alternative would be to enforce the nonpermeability condition with a penalty method (as done for Dirichlet conditions throughout this work), but this method would yield null velocity in the sharp corner of $\Gamma_{slip}$. The additional terms with the new field $l$ are
\begin{equation} \label{eq:addint}
 \int_{\Gamma_{slip}} 
 [u_r,\;u_z]\cdot[n_r,\;n_z]\; w + l\; [v_r,\;v_z]\cdot[n_r,\;n_z]\;{\mathrm{d}}s 
 +
  \int_{D} 10^{-6}\;l\;w\;r\,{\mathrm{d}}r\,{\mathrm{d}}z,
\end{equation}
with test functions $v_r,\;v_z$ of the corresponding velocity fields. Examination of \eqref{eq:addint} suggests interpreting $l$ as an additional (boundary) pressure field (pressure gradient term has a corresponding form in the weak formulation), gradient of which is an additional forcing needed to satisfy the nonpermeability constrain. The agreement between Nek5000 and FreeFem++ results obtained with free slip conditions validates our implementation. A similar discussion applies to the boundary conditions in the eigenvalue problem but, is omitted here. More details specific to the finite element discretisation can be found in \citep{pironneau1989finite}.

\section*{Appendix C : Pseudo-arclength continuation}
A solution $\bm{u}$ of a set of nonlinear algebraic equations $g_i(\bm{u}; \lambda)$ for a given parameter $\lambda$, is often computed using a Newton-Raphson algorithm. This consists of iteratively updating $\bm{u}$ by $\delta \bm{u}$, obtained by solving the linear system
\begin{equation} \label{eq:nr}
\frac{\partial g_i}{\partial u_j}(\bm u; \lambda)\,\delta  u_j=-g_i(\bm{u};\lambda).
\end{equation}
In the next iteration $\bm u^{n+1}=\bm u^n+\delta \bm u^n$, and the algorithm is continued until convergence, usually detected by a small norm of either $\bm g$ or $\delta \bm u$. If, for some value of $\lambda$, there exist a fold of the branch of $\bm u$, the Jacobian ${\partial g_i}/{\partial u_j}$ in  \eqref{eq:nr} becomes singular and the parameter $\lambda$ must be treated as an additional variable. The additional equation needed to find the new variable $\lambda$, is a pseudo-arclength condition, which constrains the solution $\bm u$ to lie in the vicinity of the known solution $\bm u_0$ identified for some $\lambda_0$, at a prescribed distance $ds= [\bm u-\bm u_0, \lambda-\lambda_0]\cdot\bm s$, where $\bm s=[\bm s_{\bm u},  s_\lambda]$ is a unit norm vector tangent to the branch and can be approximated using two previous solutions on the branch. The augmented set of equations is
\begin{equation} \label{eq:arc}
\begin{bmatrix}
\frac{\partial g_i}{\partial u_j} & \frac{\partial g_i}{\partial \lambda}\\
\bm s_{\bm u} & s_{\lambda}
\end{bmatrix}  
\begin{bmatrix}
\delta u_j \\ \delta \lambda
\end{bmatrix}
=
\begin{bmatrix}
-g_i(\bm{u},\lambda) \\ ds - [\bm u-\bm u_0, \lambda-\lambda_0]\cdot\bm s
\end{bmatrix}
\end{equation}
which is solved for $\delta \bm u$ and $\delta \lambda$, and the fields $\bm u$ and $\lambda$ are incremented by these values. The vector $\bm u$ represents any of the unknown fields constituting a solutions. It represents steady velocities and pressure in the current work but can also represent time periodic fields as in \cite{geslasubc}. Parameter $\lambda$ is usually taken to be $Re$. The solution of \eqref{eq:arc} can be computed efficiently by observing that the left hand side matrix is sparse except for the last row and the last column which yields an arrowhead matrix. Algorithms exploiting this special form exist but we found that a regular sparse solver available in UMFPACK is efficient enough for the problem sizes considered here. Upon convergence, the solution $[\bm u, \lambda]$ is mapped to $[\bm u_0, \lambda_0]$, tangent vector $\bm s$ is recomputed, and the algorithm proceeds to the next point lying on the branch. The arc length step size is controlled by $ds$ and can be adapted along the branch in order to improve efficiency. In this work, $ds$ is adapted as a function of the convergence rate of the algorithm.
In this way, more arclength points are computed around sharp turns of the branch (near folds). Various methods of arclength size adaptation and more details on numerical continuation methods can be found in \cite{allgower2012numerical}.

\bibliographystyle{apalike}

\bibliography{biblio.bib}

@article{rheingans1940power,
  title={Power swings in hydroelectric power plants},
  author={Rheingans, W.J.},
  journal={J. Fluids Eng.},
  volume={62},
  number={3},
  pages={171--177},
  year={1940},
  publisher={American Society of Mechanical Engineers Digital Collection}
}

@article{allen1924comparative,
  title={Comparative Tests on Experimental Draft-Tubes},
  author={Allen, C.M. and Winter, I.A.},
  journal={Trans. Am. Soc. Civil Eng.},
  volume={87},
  number={1},
  pages={893--926},
  year={1924},
  publisher={American Society of Civil Engineers}
}

@article{dorfler1980mathematical,
  title={Mathematical model of the pulsations in {F}rancis turbines caused by the vortex core at part load},
  author={D{\"o}rfler, P.},
  journal={Escher Wyss News;(Switzerland)},
  volume={53},
  number={1/2},
  year={1980}
}

@inproceedings{seidel2014dynamic,
  title={Dynamic loads in {F}rancis runners and their impact on fatigue life},
  author={Seidel, U. and Mende, C. and H{\"u}bner, B. and Weber, W. and Otto, A.},
  booktitle={IOP {C}onference {S}eries: {E}arth and {E}nvironmental {S}cience},
  volume={22},
  onumber={3},
  pages={032054},
  year={2014},
  organization={IOP Publishing}
}

@article{ciocan2007experimental,
  title={Experimental study and numerical simulation of the {FLINDT }draft tube rotating vortex},
  author={Ciocan, G. D. and Iliescu, M. S. and Vu, T. C. and Nennemann, B. and Avellan, F.},
  year={2007},
  journal={J. Fluids Eng.},
  volume={129},
  number={2},
  pages={146--158}
}

@article{escudier1987confined,
  title={Confined vortices in flow machinery},
  author={Escudier, M.},
  journal={Annu. Rev. Fluid Mech.},
  volume={19},
  pages={27--52},
  year={1987}
}

@article{escudier1988vortex,
  title={Vortex breakdown: observations and explanations},
  author={Escudier, M.},
  journal={Prog. Aerosp. Sci.},
  volume={25},
  number={2},
  pages={189--229},
  year={1988},
  publisher={Elsevier}
}

@article{cassidy1970observations,
  title={Observations of unsteady flow arising after vortex breakdown},
  author={Cassidy, J. J. and Falvey, H. T.},
  journal={J. Fluid Mech.},
  volume={41},
  number={4},
  pages={727--736},
  year={1970},
  publisher={Cambridge University Press}
}

@article{leibovich1978structure,
  title={The structure of vortex breakdown},
  author={Leibovich, S.},
  journal={Annu. Rev. Fluid Mech.},
  volume={10},
  pages={221--246},
  year={1978}
}

@article{benjamin1962theory,
  title={Theory of the vortex breakdown phenomenon},
  author={Benjamin, T. B.},
  journal={J. Fluid Mech.},
  volume={14},
  number={4},
  pages={593--629},
  year={1962},
  publisher={Cambridge University Press}
}

@article{zhang2009topological,
  title={Topological structure of shock induced vortex breakdown},
  author={Zhang, S. and Zhang, H. and Shu, C.-W.},
  journal={J. Fluid Mech.},
  volume={639},
  pages={343--372},
  year={2009},
  publisher={Cambridge University Press}
}

@article{faler1977disrupted,
  title={Disrupted states of vortex flow and vortex breakdown},
  author={Faler, J. and Leibovich, S.},
  journal={Phys. Fluids},
  volume={20},
  number={9},
  pages={1385--1400},
  year={1977},
  publisher={American Institute of Physics}
}

@article{billant1998experimental,
  title={Experimental study of vortex breakdown in swirling jets},
  author={Billant, P. and Chomaz, J.-M. and Huerre, P.},
  journal={J. Fluid Mech.},
  volume={376},
  pages={183--219},
  year={1998},
  publisher={Cambridge University Press}
}

@article{delbende1998absolute,
  title={Absolute/convective instabilities in the {B}atchelor vortex: a numerical study of the linear impulse response},
  author={Delbende, I. and Chomaz, J.-M. and Huerre, P.},
  journal={J. Fluid Mech.},
  volume={355},
  pages={229--254},
  year={1998},
  publisher={Cambridge University Press}
}

@article{gallaire2006spiral,
  title={Spiral vortex breakdown as a global mode},
  author={Gallaire, F. and Ruith, M. and Meiburg, E. and Chomaz, J.-M. and Huerre, P.},
  journal={J. Fluid Mech.},
  volume={549},
  pages={71--80},
  year={2006},
  publisher={Cambridge University Press}
}

@article{ruith2003three,
  title={Three-dimensional vortex breakdown in swirling jets and wakes: direct numerical simulation},
  author={Ruith, M. R. and Chen, P. and Meiburg, E. and Maxworthy, T.},
 journal={J. Fluid Mech.},
  volume={486},
  pages={331--378},
  year={2003},
  publisher={Cambridge University Press}
}

@article{vyazmina2009bifurcation,
  title={The bifurcation structure of viscous steady axisymmetric vortex breakdown with open lateral boundaries},
  author={Vyazmina, E. and Nichols, J. W. and Chomaz, J.-M. and Schmid, P. J.},
  journal={Phys. Fluids},
  volume={21},
  number={7},
  year={2009},
  publisher={AIP Publishing}
}

@article{gallaire2003mode,
  title={Mode selection in swirling jet experiments: a linear stability analysis},
  author={Gallaire, F. and Chomaz, J.-M.},
  journal={J. Fluid Mech.},
  volume={494},
  pages={223--253},
  year={2003},
  publisher={Cambridge University Press}
}

@article{tsitverblit1993vortex,
  title={Vortex breakdown in a cylindrical container in the light of continuation of a steady solution},
  author={Tsitverblit, N.},
  journal={Fluid Dyn. Res.},
  volume={11},
  number={1-2},
  pages={19},
  year={1993},
  publisher={IOP Publishing}
}

@article{chen1995bifurcation,
  title={Bifurcation for flow past a cylinder between parallel planes},
  author={Chen, J.-H. and Pritchard, W. G. and Tavener, S. J.},
  journal={J. Fluid Mech.},
  volume={284},
  pages={23--41},
  year={1995},
  publisher={Cambridge University Press}
}

@article{oberleithner2011three,
  title={Three-dimensional coherent structures in a swirling jet undergoing vortex breakdown: stability analysis and empirical mode construction},
  author={Oberleithner, K. and Sieber, M. and Nayeri, C. N. and Paschereit, C. O. and Petz, C. and Hege, H.-C. and Noack, B. R. and Wygnanski, I.},
  journal={J. Fluid Mech.},
  volume={679},
  pages={383--414},
  year={2011},
  publisher={Cambridge University Press}
}

@article{liang2005experimental,
  title={An experimental investigation of swirling jets},
  author={Liang, H. and Maxworthy, T.},
  journal={J. Fluid Mech.},
  volume={525},
  pages={115--159},
  year={2005},
  publisher={Cambridge University Press}
}

@article{healey2008inviscid,
  title={Inviscid axisymmetric absolute instability of swirling jets},
  author={Healey, J. J.},
  journal={J. Fluid Mech.},
  volume={613},
  pages={1--33},
  year={2008},
  publisher={Cambridge University Press}
}

@article{lopez1994bifurcation,
  title={On the bifurcation structure of axisymmetric vortex breakdown in a constricted pipe},
  author={Lopez, J. M.},
  journal={Phys. Fluids},
  volume={6},
  number={11},
  pages={3683--3693},
  year={1994},
  publisher={American Institute of Physics}
}

@article{meliga2011control,
  title={Control of axisymmetric vortex breakdown in a constricted pipe: nonlinear steady states and weakly nonlinear asymptotic expansions},
  author={Meliga, P. and Gallaire, F.},
  journal={Phys. Fluids},
  volume={23},
  number={8},
  year={2011},
  publisher={AIP Publishing}
}

@article{shtern1999collapse,
  title={Collapse, symmetry breaking, and hysteresis in swirling flows},
  author={Shtern, V. and Hussain, F.},
  journal={Annu. Rev. Fluid Mech.},
  volume={31},
  number={1},
  pages={537--566},
  year={1999},
  publisher={Annual Reviews 4139 El Camino Way, PO Box 10139, Palo Alto, CA 94303-0139, USA}
}

@inproceedings{muller2021prediction,
  title={Prediction of vortex precession in the draft tube of a model hydro turbine using mean field stability theory and stochastic modelling},
  author={M{\"u}ller, J. S. and Sieber, M. and Litvinov, I. and Shtork, S. and Alekseenko, S. and Oberleithner, K.},
  booktitle={IOP Conference Series: Earth and Environmental Science},
  volume={774},
  number={1},
  pages={012003},
  year={2021},
  organization={IOP Publishing}
}

@article{pasche2017part,
  title={Part load vortex rope as a global unstable mode},
  author={Pasche, S. and Avellan, F. and Gallaire, F.},
  journal={J. Fluids Eng.},
  volume={139},
  number={5},
  pages={051102},
  year={2017},
  publisher={American Society of Mechanical Engineers}
}

@article{susan2010analysis,
  title={Analysis and prevention of vortex breakdown in the simplified discharge cone of a {F}rancis turbine},
  author={Susan-Resiga, R. and Muntean, S. and Hasmatuchi, V. and Anton, I. and Avellan, F.},
  year={2010},
  journal={J. Fluids Eng.},
  volume={132},
  number={5}
}

@article{mantivc2014self,
  title={Self-consistent mean flow description of the nonlinear saturation of the vortex shedding in the cylinder wake},
  author={Manti{\v{c}}-Lugo, V. and Arratia, C. and Gallaire, F.},
  journal={Phys. Rev. Lett.},
  volume={113},
  number={8},
  pages={084501},
  year={2014},
  publisher={APS}
}

@article{sarpkaya1974effect,
  title={Effect of the adverse pressure gradient on vortex breakdown},
  author={Sarpkaya, T.},
  journal={AIAA J.},
  volume={12},
  number={5},
  pages={602--607},
  year={1974}
}

@article{susan2006analysis,
  title={Analysis of the swirling flow downstream a {F}rancis turbine runner},
  author={Susan-Resiga, R. and Dan Ciocan, G. and Anton, I. and Avellan, F.},
  year={2006},
  journal={J. Fluids Eng.},
  volume={128},
  number={1}
}

@article{freefem,
  AUTHOR = {Hecht, F.},
  TITLE = {New development in {F}reeFem++},
  JOURNAL = {J. Numer. Math.},
  FJOURNAL = {Journal of Numerical Mathematics},
  VOLUME = {20}, YEAR = {2012},
  NUMBER = {3-4}, PAGES = {251--265},
  ISSN = {1570-2820},
  MRCLASS = {65Y15},
  MRNUMBER = {3043640},
  URL = {https://freefem.org/}
}

@Misc{nek5000,
  Author = "Fischer, P. F. and Lottes, J. W. and Kerkemeier, S. G.",
  Title  = "{nek5000} {W}eb page",
  Note   = "http://nek5000.mcs.anl.gov",
  Year   = "2008"}

@book{pironneau1989finite,
  title={Finite Element Methods for Fluids},
  author={Pironneau, O.},
  year={1989},
  publisher={Wiley Chichester}
}

@article{amini2023upper,
  title={Upper part-load instability in a reduced-scale {F}rancis turbine: an experimental study},
  author={Amini, A. and Vagnoni, E. and Favrel, A. and Yamaishi, K. and M{\"u}ller, A. and Avellan, F.},
  journal={Exp. Fluids},
  volume={64},
  number={6},
  pages={110},
  year={2023},
  publisher={Springer}
}

@article{barkley2006linear,
  title={Linear analysis of the cylinder wake mean flow},
  author={Barkley, D.},
  journal={Europhys. Lett.},
  volume={75},
  number={5},
  pages={750},
  year={2006},
  publisher={IOP Publishing}
}

@article{xin2001linear,
  title={Linear stability analyses of natural convection flows in a differentially heated square cavity with conducting horizontal walls},
  author={Xin, S. and Le Qu{\'e}r{\'e}, P.},
  journal={Phys. Fluids},
  volume={13},
  number={9},
  pages={2529--2542},
  year={2001},
  publisher={American Institute of Physics}
}

@inproceedings{susan2009three,
  title={Three-dimensional versus two-dimensional axisymmetric analysis for decelerated swirling flows},
  author={Susan-Resiga, R. F. and Muntean, S. and Tanasa, C. and Bosioc, A.},
  booktitle={Proceedings of the 13th International Conference on Fluid Flow, Budapest, Hungary},
  pages={862--869},
  year={2009}
}

@phdthesis{viola2016resonance,
  title={Resonance in swirling wakes and sloshing waves: non-normal and sublinear effects},
  author={Viola, F.},
  year={2016},
  school={Ecole Polytechnique F{\'e}d{\'e}rale de Lausanne}
}

@book{allgower2012numerical,
  title={Numerical {C}ontinuation {M}ethods: {A}n {I}ntroduction},
  author={Allgower, E. L. and Georg, K.},
  volume={13},
  year={2012},
  address={Cham},
  publisher={Springer Science \& Business Media}
}

@article{Jeong_Hussain_1995, 
title={On the identification of a vortex}, volume={285}, DOI={10.1017/S0022112095000462}, journal={J. Fluid Mech.},
 author={Jeong, J. and Hussain, F.}, year={1995}, pages={69–94}
 }

@article{avila2013streamwise,
  title={Streamwise-localized solutions at the onset of turbulence in pipe flow},
  author={Avila, M. and Mellibovsky, F. and Roland, N. and Hof, B.},
  journal={Phys. Rev. Lett.},
  volume={110},
  number={22},
  pages={224502},
  year={2013},
  publisher={APS}
}

@article{zheng2025ghost,
  title={Ghost states underlying spatial and temporal patterns: {H}ow nonexistent invariant solutions control nonlinear dynamics},
  author={Zheng, Z. and Beck, P. and Yang, T. and Ashtari, O. and Parker, J. P. and Schneider, T. M.},
  journal={Phys. Rev. E},
  volume={112},
  number={2},
  pages={024212},
  year={2025},
  publisher={APS}
}

@book{strogatz2001nonlinear,
  title={Nonlinear Dynamics and Chaos: With Applications to Physics, Biology, Chemistry and Engineering},
  author={Strogatz, S. H.},
  volume={1},
  year={2001},
  publisher={Westview press}
}

@article{wang1997effect,
  title={The effect of slight viscosity on a near-critical swirling flow in a pipe},
  author={Wang, S. and Rusak, Z.},
  journal={Phys. Fluids},
  volume={9},
  number={7},
  pages={1914--1927},
  year={1997},
  publisher={American Institute of Physics}
}

@phdthesis{elenavphd,
  title={Bifurcations in a swirling flow},
  author={Vyazmina, E.},
  year={2010},
  school={Ecole Polytechnique}
}

@book{charru2011hydrodynamic,
  title={Hydrodynamic Instabilities},
  author={Charru, F.},
  volume={37},
  year={2011},
  publisher={Cambridge University Press}
}

@misc{suppmat,
  note = "See Supplemental Material at
    URL-will-be-inserted-by-publisher/.mp4"
}

@article{Meliga_Gallaire_Chomaz_2012, 
title={A weakly nonlinear mechanism for mode selection in swirling jets}, volume={699}, DOI={10.1017/jfm.2012.93}, journal={J. Fluid Mech.}, author={Meliga, P. and Gallaire, F. and Chomaz, J.-M.}, year={2012}, pages={216–262}
}

@article{ruith2002,
    author = {Ruith, M. R. and Meiburg, E.},
    title = {Breakdown Modes of Swirling Jets with Coflow},
    journal = {Phys. Fluids},
    volume = {14},
    number = {9},
    pages = {S11-S11},
    year = {2002},
    month = {09},
    issn = {1070-6631},
    doi = {10.1063/1.4739201},
    url = {https://doi.org/10.1063/1.4739201},
    eprint = {https://pubs.aip.org/aip/pof/article-pdf/14/9/S11/19159144/s11_1_online.pdf},
}

@book{ern2004theory,
  title={Theory and Practice of Finite Elements},
  author={Ern, A. and Guermond, J.-L.},
  volume={159},
  year={2004},
  publisher={Springer}
}

@article{Gesla24,
 title={On the origin of circular rolls in rotor-stator flow},
  volume={1000},
   DOI={10.1017/jfm.2024.1011},
    journal={J. Fluid Mech.},
     author={Gesla, A. and Duguet, Y. and Le Quéré, P. and Martin Witkowski, L.},
      year={2024},
       pages={A47}
       }

@article{Cerqueira_Sipp_2014,
        title={Eigenvalue sensitivity, singular values and discrete frequency selection mechanism in noise amplifiers: the case of flow induced by radial wall injection}, 
        volume={757}, 
        DOI={10.1017/jfm.2014.513},
         journal={J. Fluid Mech.},
          author={Cerqueira, S. and Sipp, D.}, year={2014}, pages={770–799}
         }

@article{benjamin1978bifurcation,
  title={Bifurcation phenomena in steady flows of a viscous fluid. {I}. {T}heory},
  author={Benjamin, T. B.},
  journal={Proc. R. Soc. Lond., Ser. A },
  volume={359},
  number={1696},
  pages={1--26},
  year={1978},
  publisher={The Royal Society London}
}

@article{geslasubc,
  title = {Subcritical axisymmetric solutions in rotor-stator flow},
  author = {Gesla, A. and Duguet, Y. and Le Qu\'er\'e, P. and Martin Witkowski, L.},
  journal = {Phys. Rev. Fluids},
  volume = {9},
  issue = {8},
  pages = {083903},
  numpages = {28},
  year = {2024},
  month = {Aug},
  publisher = {American Physical Society},
  doi = {10.1103/PhysRevFluids.9.083903},
  url = {https://link.aps.org/doi/10.1103/PhysRevFluids.9.083903}
}

@article{Yim19SelfConsitent,
  title={Self-consistent triple decomposition of the turbulent flow over a backward facing step under finite amplitude harmonic forcing},
  author={Yim, E. and Meliga, P. and Gallaire, F.},
  journal={Proc. Roy. Soc. Lond. {\em A}},
  volume={475},
  pages={20190018},
  year={2019},
}

@article{head_paper_1,
        title={Linear stability and structural sensitivity of
a swirling jet in a {F}rancis turbine draft tube}, 
        volume={}, 
        DOI={},
         journal={under review in J. Fluid Mech.},
          author={Toledo, L. C. and Gesla, Artur and Yim, E.}, year={2026}, pages={}
         }

@article{openpipeflow,
               title={The {O}penpipeflow {N}avier--{S}tokes Solver},
               author={Willis, A. P.},
               journal={SoftwareX},
               volume={6},
               pages={124--127},
               year={2017},
               DOI = {https://doi.org/10.1016/j.softx.2017.05.003 },
               url = {http://arxiv.org/abs/1705.03838 }
            }

@article{wangPof1996a,
    author = {Wang, S. and Rusak, Z.},
    title = {On the stability of an axisymmetric rotating flow in a pipe},
    journal = {Phys. Fluids},
    volume = {8},
    number = {4},
    pages = {1007-1016},
    year = {1996},
    month = {04},
    issn = {1070-6631},
    doi = {10.1063/1.868882},
    url = {https://doi.org/10.1063/1.868882},
    eprint = {https://pubs.aip.org/aip/pof/article-pdf/8/4/1007/19244485/1007_1_online.pdf}
}

@article{Wang_Rusak_Gong_Liu_2016, title={On the three-dimensional stability of a solid-body rotation flow in a finite-length rotating pipe}, volume={797}, DOI={10.1017/jfm.2016.223}, journal={J. Fluid Mech.}, author={Wang, S. and Rusak, Z. and Gong, R. and Liu, F.}, year={2016}, pages={284–321}
}

@article{Feng_Liu_Rusak_Wang_2018, title={Dynamics of a perturbed solid-body rotation flow in a finite-length straight rotating pipe}, volume={846}, DOI={10.1017/jfm.2018.245}, journal={J. Fluid Mech.}, author={Feng, C. and Liu, F. and Rusak, Z. and Wang, S.}, year={2018}, pages={1114–1152}
}

@article{Qiao_Shi_Meng_Wang_Liu_2025, title={Instability modes of the {L}amb–{O}seen vortex within a finite-length pipe}, volume={1019}, DOI={10.1017/jfm.2025.10595}, journal={J. Fluid Mech.}, author={Qiao, Y. and Shi, Y. and Meng, X. and Wang, S. and Liu, F.}, year={2025}, pages={A47}
}

@article{QiaoPof2025,
    author = {Qiao, Y. and Shi, Y. and Wang, S. and Liu, F.},
    title = {Bifurcation paths and stability of non-unique solutions of the axisymmetric {L}amb–{O}seen vortex in a pipe of finite-length},
    journal = {Phys. Fluids},
    volume = {37},
    number = {6},
    pages = {064105},
    year = {2025},
    month = {06},
    issn = {1070-6631},
    doi = {10.1063/5.0275304},
    url = {https://doi.org/10.1063/5.0275304},
    eprint = {https://pubs.aip.org/aip/pof/article-pdf/doi/10.1063/5.0275304/20544523/064105_1_5.0275304.pdf},
}

@article{wang1997dynamics,
  title={The dynamics of a swirling flow in a pipe and transition to axisymmetric vortex breakdown},
  author={Wang, S. and Rusak, Z.},
  journal={J. Fluid Mech.},
  volume={340},
  pages={177--223},
  year={1997},
  publisher={Cambridge University Press}
}

@article{feng2017energy,
  title={The energy transfer mechanism of a perturbed solid-body rotation flow in a rotating pipe},
  author={Feng, C. and Liu, F. and Rusak, Z. and Wang, S.},
  journal={Acta Mech. Sin.},
  volume={33},
  number={2},
  pages={274--283},
  year={2017},
  publisher={Springer}
}

@phdthesis{Muri02, 
address={Lausanne},
 title={Numerical simulation and flow analysis of an elbow diffuser},
   DOI={10.5075/epfl-thesis-2527},
 school={EPFL},
  author={Mauri, S.}, 
  year={2002}
 }

\end{document}